\documentclass[aps,showpacs,showkeys,eqsecnum]{revtex4}
\usepackage{amsfonts,amsmath,amssymb,bm,graphicx}

\newcommand{\gs}{g_S}
\newcommand{\gw}{g_{(1)}}
\newcommand{\gv}{g_{V}}
\newcommand{\gva}{g_{\textit VA}}
\newcommand{\gsv}{g_{(3/2)}}
\newcommand{\veg}{\mathbf}
\newcommand{\charge}{\hskip 0.02cm {e} \hskip0.02cm}
\newcommand{\mmt}{\veg{p}}

\def\l{\left}

\newcommand{\aeq}{\begin{equation}}
\newcommand{\ceq}{\end{equation}}
\newcommand{\aec}{\begin{eqnarray}}
\newcommand{\cec}{\end{eqnarray}}
\newcommand{\ase}{\begin{subequations}}
\newcommand{\cse}{\end{subequations}}

\renewcommand{\(}{\left(}
\renewcommand{\)}{\right)}
\renewcommand{\[}{\left[}
\renewcommand{\]}{\right]}

\newcommand{\Tr}{\textup{Tr}}

\newcommand{\diracrep}{\left(1/2,0\right)\oplus\left(0,1/2\right)}

\newcommand{\tmrep}{\left(3/2,0\right)\oplus\left(0,3/2\right)}
\newcommand{\procarep}{\left(1/2,1/2\right)}
\newcommand{\rsrep}{\left(1/2,1/2\right)\otimes[\left(1/2,0\right)\oplus\left(0,1/2\right)]}

\newcommand{\lag}{\mathcal{L}}

\newcommand{\M}{\mathcal{M}}

\newcommand{\free}{\mbox{\footnotesize free}}
\newcommand{\inte}{\mbox{\footnotesize int}}
\newcommand{\ket}{\rangle}
\newcommand{\bra}{\langle}

\renewcommand{\a}{\alpha}
\renewcommand{\b}{\beta}
\newcommand{\m}{\mu}
\newcommand{\n}{\nu}
\renewcommand{\o}{\omega}
\newcommand{\g}{\gamma}
\renewcommand{\d}{\delta}
\newcommand{\h}{\eta}
\newcommand{\z}{\zeta}
\newcommand{\f}{\phi}

\newcommand{\y}{\psi}
\renewcommand{\l}{\lambda}
\newcommand{\s}{\sigma}

\newcommand{\q}{\theta}
\newcommand{\e}{\epsilon}
\newcommand{\x}{\xi}

\newcommand{\p}{\pi}

\newcommand{\G}{\Gamma}
\renewcommand{\P}{\Pi}

\newcommand{\pd}{\partial}

\DeclareFontFamily{OT1}{pzc}{}
\DeclareFontShape{OT1}{pzc}{m}{it}{<-> s * [1.10] pzcmi7t}{}
\DeclareMathAlphabet{\mathpzc}{OT1}{pzc}{m}{it}

\begin{document}

\date{\today }

\title{ Electromagnetic multipole moments of elementary spin-1/2, 1, and 3/2 particles}

\author{E. G. Delgado-Acosta}\email{german@ifisica.uaslp.mx}
\affiliation{Instituto de F\'{\i}sica, Universidad Aut\'onoma de San Luis Potos\'{\i},
Av. Manuel Nava 6, San Luis Potos\'{\i}, S.L.P. 78290, M\'exico}

\author{M. Kirchbach}\email{mariana@ifisica.uaslp.mx}
\affiliation{Instituto de F\'{\i}sica, Universidad Aut\'onoma de San Luis Potos\'{\i},
Av. Manuel Nava 6, San Luis Potos\'{\i}, S.L.P. 78290, M\'exico}

\author{M. Napsuciale}\email{mauro@fisica.ugto.mx}
\affiliation{Departamento de F\'{\i}sica, Universidad de Guanajuato,
Lomas del Campestre 103, Fraccionamiento Lomas del Campestre, Le\'on, Guanajuato, 37150, M\'exico}
\email{mauro@fisica.ugto.mx}

\author{S. Rodr\'{\i}guez}\email{simonrodriguez@uadec.edu.mx}
\affiliation{Facultad de Ciencias F\'{\i}sico Matem\'aticas, Universidad Aut\'onoma de Coahuila,
Edificio ``D", Unidad Camporredondo, Saltillo, Coahuila, 25280, M\'exico}

\begin{abstract}{
We study multipole decompositions of the electromagnetic currents of spin-1/2, 1, and 3/2 particles described in terms
of representation specific wave equations which are second order in the momenta and which emerge within the
recently elaborated Poincar\'e covariant projector method,
where the respective Lagrangians explicitly depend on the Lorentz group generators of the
representations of interest.  The currents are then the ordinary linear Noether currents related to phase invariance, and
present themselves always as two-terms motion-- plus spin-magnetization currents. The spin-magnetization currents appear
weighted by the gyromagnetic ratio, $g$, a free parameter in the method  which we fix either by unitarity of
forward Compton scattering amplitudes in the ultraviolet for spin-1,  and spin-3/2,
or, in the spin-1/2 case, by their asymptotic vanishing, thus ending up in all three cases with the universal $g$ value of $g=2$.
Within the method under discussion we calculate the electric multipoles of the above spins for the spinor--, the four-vector,
and the four-vector--spinor representations, and find it favorable in some aspects specifically in comparison with
the conventional  Proca-, and Rarita-Schwinger frameworks.
We furthermore  attend to the most general non-Lagrangian spin-3/2 currents which are allowed by Lorentz invariance
to be up to third order in the momenta and construct the linear-current equivalent of identical multipole moments of one of them.
We conclude that non-linear non-Lagrangian spin-3/2 currents are not necessarily more general and more advantageous than the linear spin-3/2  Lagrangian current
emerging within the covariant projector formalism.
Finally,  we test the representation dependence of the multipoles by placing spin-1 and spin-3/2  in the respective
(1,0)$\oplus$(0,1), and (3/2,0)$\oplus$(0,3/2) single-spin representations. We observe representation independence of the
charge monopoles and the magnetic dipoles, in contrast to the higher multipoles, which turn out to be representation dependent.
In particular, we find the bi-vector $(1,0)\oplus (0,1)$ to be  characterized by
an electric quadrupole moment of opposite sign to the one found in $(1/2,1/2)$, and consequently, to the $W$ boson.
This observation allows to explain the positive electric quadrupole moment  of the $\rho$ meson extracted from recent
analyzes of the $\rho$ meson electric form factor. Our finding  points toward the possibility that the $\rho$ meson could
transform as part of  an anti-symmetric tensor with an $a_{1}$ meson-like state as its representation companion,
a possibility consistent with the empirically established $\rho$--, and $a_{1}$ vector meson dominance of the hadronic
vector---,  and axial-vector currents.
}
\end{abstract}

%03.65.Pm Relativistic wave equations
%Electromagnetic moments, nuclear, 21.10.Ky
%Electromagnetic interactions, 13.40.-f
%13.40.Em Electric and magnetic moments

\maketitle
%-------------------------------------------------------------------------------------------------
\tableofcontents

\section{Introduction}

%-------------------------------------------------------------------------------------------------
The electromagnetic characteristics of elementary particles continue being one of the key issues in contemporary physics
research, both experimental  and theoretical. The reason is that a great deal of our  knowledge on the fundamental
 properties of matter is in first instance obtained from theoretical analyzes of measured cross sections of
 electromagnetic processes such as nuclear reactions
induced by real photons, or elastic and
inelastic electron scattering off proton or nuclear targets. In this type of processes, particles with spins higher than 1/2 can arise
both as intermediate virtual, or, real outgoing states. As an example we wish to mention the reaction of a quasi-free knock-out
of a $\Delta (1232)$ particle in the inelastic electron scattering on $^3He$ \cite{Schulze:1992mb}. The evaluation of this process relies
upon the electromagnetic moments of the $\Delta (1232)$ particle, a spin-3/2 state. The electromagnetic spin-3/2  current has
been  widely analyzed, predominantly within the Rarita-Schwinger framework \cite{Lorce:2009br} where a spin-$J$ fermion is
considered as the highest spin, $J=K+\frac{1}{2}$, in the totally symmetric rank-$K$ tensor-spinor, $\psi_{\mu_1...\mu_K}$.
As is well known, the Rarita-Schwinger Lagrangian is linear in the momenta and takes into account certain auxiliary conditions,
supposed to restrict the tensor-spinor space to the desired $2(2J+1)$ fermion--anti-fermion degrees of freedom. Although widely
used, the Rarita-Schwinger framework is known to suffer several pathologies, among them, the acausal propagation of the classical
wave fronts of the particle within an electromagnetic environment. Recently, a second order  formalism for  the description
of the electromagnetic interactions of particles with spins in terms of $\psi_{\mu_1...\mu_K}$ has been developed in
\cite{Napsuciale:2006wr} which was shown to be free from the problem of acausal propagation for $K=1$, provided the spin-3/2
gyromagnetic ratio, $\gsv$, were to take the universal value of $\gsv$. Within the latter formalism,
which naturally incorporates the necessary auxiliary conditions, the $J=K+ \frac{1}{2}$ sector of $\psi_{\mu_1...\mu_K}$ is pinned down by a
covariant projector constructed from the two Casimir operators of the Poincar\'e group, the squared  linear momentum, $P^2$, and
the squared  Pauli-Lubanski vector, $W^2$. The formalism is applicable to boson fields as well and has been employed in the
calculation of the Compton scattering  cross section off vector particles \cite{Napsuciale:2007ry}. In the latter  work, the case
has been made that while Poincar\'e invariance prescribes  the cross section under investigation to depend on only two parameters,
in turn identified as the spin-1 gyromagnetic factor, $\gw$, and the coupling $\x$ associated to parity violating Lorentz structures,
it is insufficient to fix their values. To fix those values, additional dynamical requirements need to be imposed on the cross sections.
In particular,
in imposing the unitarity condition on the Compton scattering amplitudes in the ultraviolet, the $\xi=0$ value, and, again, the
universal  $\gw=2$ value have been encountered in \cite{Napsuciale:2007ry}.

The same path has been pursued  in \cite{DelgadoAcosta:2009ic} with the aim to evaluate the Compton scattering process off a
spin $3/2$ target.  Confining to parity conservation, the amplitudes have been found again to  depend   on  $g$  alone. Moreover,
for $\gsv=2$, the forward differential cross  section was shown to become finite in accord with unitarity. In this way, an
independent confirmation of the causality argument for $g=2$ universality has been obtained. Knowing the value of the gyromagnetic ratio,  provides a reliable
basis for the calculation of the multipole moments of particles with spins ranging for  1/2 to 3/2.

\begin{quote}
The goal of the present work is to find  within the Poincar\'e covariant second order projector formalism  expressions for the
electromagnetic multipole moments of particles with spins 1/2, 1, and 3/2, exploring both single-spin--, and multiple-spin
representations of the Lorentz group.
\end{quote}

In the literature, the electromagnetic multipole  moments of a fundamental Dirac particle have been well understood
\cite{PhysRev.87.688,Aitchison:2003tq} and reproducing their values is  mandatory for establishing credibility of any new high spin
method, including the present one. Below we  show that the Poincar\'e covariant second order formalism yields the same
multipoles for a spin-1/2 particle  as the Dirac theory. As to the higher spins 1 and 3/2, and in comparison
to the Proca and Rarita-Schwinger frameworks, the formalism under investigation will be shown to bring some notable
improvements in our understanding of the electromagnetic properties of the above elementary particles.

To be specific,  the electromagnetic current of a spin 3/2 particle is often designed in the literature by
decomposing it into the basis of the most general tensors compatible with Lorentz invariance, not necessarily
restricted to first order in the momenta \cite{glaser,gourdin1,gourdin2,Nozawa:1990gt,Lorce:2009br}. The
Poincar\'e covariant projector method instead yields a current expressed in terms of the generators of the Poincar\'e  group for
the representation of interest which is linear in the momenta and generates a full flashed
spin-3/2 contribution to the electric quadrupole moment of the particle under discussion, in contrast to the
Rarita-Schwinger formalism which, as we will show, incorporates only the contribution of the corresponding spin-1/2 sector.

The paper is structured as follows. In the next section we review the Poincar\'e covariant second order projector
formalism for the spinor and vector-spinor representations  and elaborate the formalism
for the $(1,0)\oplus (0,1)$ (bi-vector) and $(3/2,0)\oplus (0,3/2)$ representations with the emphasis on the
emerging electromagnetic currents. Special attention is paid to the development of the formalism for the bi-vector
representation in the form of an
anti-symmetric second rank tensor. Section III summarizes
the definitions of the electromagnetic multipole moments to be used in the paper. In section IV  we  present  a side by side
comparison of the electromagnetic multipole moments of spin-1/2, 1, and 3/2 particles following from the respective Dirac, Proca and Rarita-Schwinger frameworks, on the one side,  with same observables following from the Poincar\'e covariant second order formalism,
on the other. In section V we  discuss and a summarize  our results. The paper closes with concise conclusions and has one Appendix.

%-------------------------------------------------------------------------------------------------

\section{Covariant projector formalism and electromagnetic  currents}

%-------------------------------------------------------------------------------------------------

The idea underlying the Poincar\'e covariant second order projector formalism is that a state, $\y^{(m,s)}$,  of mass $m$, and
spin $s$ at rest, residing within a  given representation of the Poincar\'e group, can be pinned down unambiguously in any inertial
frame by means of an appropriately designed covariant projector, ${\mathcal P}^{(m,s)}$. The latter expresses
in terms of the two Casimir operators of the group, the squared momentum operator, $P^2$, and the squared Pauli-Lubanski
operator, $W^2$. For illustrative purposes we here recall the form of such a projector for the $(1/2,1/2)\otimes [(1/2,0)\oplus (0,1/2)]$  representation (four-vector--spinor representation in the following) whose wave function, $\psi_\mu$, has 16 degrees of freedom distributed
over one spin-3/2, and two  spin-1/2 fermions of opposite parities. In this case, one finds \cite{Napsuciale:2006wr},
\begin{eqnarray}
{\mathcal P}^{\left(m,\frac{3}{2}\right)}\y^{\left(m,\frac{3}{2}\right)}\, &=&\y^{\left(m,\frac{3}{2}\right)},\nonumber\\
{\mathcal P}^{\left(m,\frac{3}{2}\right)}&=& -\frac{1}{3}
\left( \frac{W^2}{m^2}+\frac{3}{2}\left( \frac{3}{2} - 1\right)\frac{P^2}{m^2}\veg{1}_{16\times 16}\right),\label{Gl1}
\end{eqnarray}
where $\veg{1}_{16\times 16}$ stands for the unit matrix in the  16-dimensional vector-spinor representation. Notice that
${\mathcal P}^{\left(m,\frac{3}{2}\right)}$ is a $16\times 16$ dimensional four- vector-spinor object which carries  two Lorentz
indices, as usually denoted by lowercase Greek letters,  and two Dirac-spinor labels, to be here denoted by capital Latin letters. Subsequently,
as explained in  \cite{Napsuciale:2006wr}, the wave
equation for $\psi_{\mu}$ can be cast in the most general covariant form according to,
\aeq(-\G_{AB \alpha \eta ; \m\n}\partial^\m \partial^\n-m^2 \delta_{AB}g_{\alpha\eta})\y^{(m,\frac{3}{2})}_B\, ^\eta =0,
\label{gamma_tensor}
\ceq
and in terms of the Lorentz tensor $\G_{AB\alpha\eta;\mu \nu}$,  carrying four covariant--, and two spinor indices.
In the labeling of the $\G$ tensor we placed a semi-colon as a demarcation sign  between the Lorentz indices of the tensor
which contract with the Lorentz indices of the four-vector representation, and those which contract with the two derivatives.
The explicit form of  $\G_{AB\alpha \eta ; \m\n}$  has been worked out in \cite{Napsuciale:2006wr} and will be presented again in
due place below. For the time being, suffice to recall that  $\G_{AB \alpha \eta ; \m\n}$  can be split into a  symmetric (SYM)
and an anti-symmetric (AS) tensor, according to
\begin{eqnarray}
\G_{AB\alpha \eta ; \m\n}=\G^{\textup{ SYM}}_{AB\alpha \eta ; \m\n} &+&\G^{\textup{AS}}_{AB\alpha \eta;  \m\n},\nonumber\\
\G_{AB\alpha \eta ; \m\n}^{\textup{SYM}} = \frac{1}{2}\left(
\G_{AB\alpha \eta ; \m\n}+ \G_{AB \alpha \eta ; \n \m}  \right), &\quad&
\G_{AB\alpha \eta ; \m\n}^{\textup{AS}} =
\frac{1}{2}\left(
\G_{AB\alpha \eta ; \m\n}- \G_{AB \alpha \eta ;\n \m} \right),
\label{SYM_ASYM}
\end{eqnarray}
and that in the absence of interactions, the  $\G_{AB\alpha \eta ;  \m\n}^{\textup{ AS}}$ term does not provide any contribution.
In order to verify  this, we drop the Dirac spinor indices, and the mass and spin labels as well, in which case (\ref{gamma_tensor}) becomes more transparent,
\aeq
(-\G_{\alpha \eta ;\m\n}\pd^\m \pd^\n -m^2\delta_{\alpha\eta})\y^\eta =0.
\ceq
Because of the commutativity of  $\pd^\m$ and $\pd^\n$, it is obvious, that for free particles  contributions of the type
 $\G^{\textup{AS}}_{\alpha \eta; \m\n}\pd^\m\pd^\n$  nullify.
This contrasts the situation in which the  particles are interacting via a gauge field, $\pd^\mu\rightarrow D^\mu=\pd^\mu+i \charge A^\mu$,
in which case  the covariant derivatives no longer  commute,
\aeq
[D^\m,D^\n]=i \charge F^{\m\n}.
\ceq
One of the essential advantages of the Poincar\'e covariant projector formalism over Rarita-Schwinger's framework is
that within the former  the anti-symmetric part of  $\G_{\alpha\eta; \m\n}$  can be designed
in the most  general way compatible with Poincar\'e invariance,
thus guarantying completeness of the equation.
Subsequently, we will occasionally refer to the Poincar\'e covariant
second order projector formalism developed in  \cite{Napsuciale:2006wr} as NKR formalism, for brevity. Towards our goal,
the description of the electromagnetic multipole moments of particles with spin-1/2, spin-1, and spin-3/2, we  present below
expressions for the respective electromagnetic currents as they emerge within the NKR formalism.
%---------------------------------------------------------------------------------------------------------------------------------------
\subsection{The spin-1/2 current in the fundamental
$\left(\frac{1}{2},0\right)\oplus \left(0, \frac{1}{2} \right)$ spinor representation}
%---------------------------------------------------------------------------------------------------------------------------------------
The NKR formalism for the fundamental  spinor $(S)$ representation, $\diracrep$, was addressed in \cite{DelgadoAcosta:2010nx} and the one
loop structure was studied in \cite{AngelesMartinez:2011nt}. The $\Gamma$
tensor corresponding to the one defined in (\ref{gamma_tensor}) carries only two Lorentz indices and the related equation of
motion reads,
\aeq
(-\G^S_{\m\n}\partial^\m \partial ^\n-m^2)\y^{\( m,\frac{1}{2}\)}=0,
\ceq
%where we suppressed the mass and spin-labels of the field for the sake of simplifying  notations. 
in the following we supress the mass and spin-labels of the field for the sake of simplifying  notations.
In the present work we restrict to  parity conserving processes, in which
case the $\Gamma^S_{\mu\nu}$ decomposition simplifies to,
\aeq
\G^S_{\m\n}=g_{\m\n}-i\gs M^S_{\m\n}.
\label{g_Sir}
\ceq
Here, $M^S_{\m\n}$ are the Lorentz group generators in the $\diracrep$ representation, and are given by,
\aeq
M^S_{\m\n}=\frac{1}{2}\s_{\m\n}, \quad \sigma_{\mu\nu}=\frac{i}{2}\left[ \gamma_\mu,\gamma_\nu\right],
\label{Dirac_genr}
\ceq
where $\gamma_\mu$ stand for the conventional Dirac matrices. Upon introducing  small Latin letters, $i,j=1,2,3$, to denote as usual the space-like Lorentz indices,
the $M_{ij}^S$ and $M^S_{0i}$ generators become in their turn the well known
pseudo-vectorial and vectorial generators of rotations $\veg{J}$, and boosts, $\veg{K}$,  according to
\begin{align}\label{spinorgen}
(M^S)^{0i}=K_{i}%=-\frac{\sigma_{i}}{2}, 
,&\qquad
(M^S)^{ij}=\e_{ijk}J_{k}.
%=\e_{ijk}\frac{i\sigma_k}{2},
\end{align}
%where $\sigma_k$ denote the standard Pauli matrices.
The associated free Lagrangian is then obtained as
\aeq
\lag^S_{\textup{ free}} =(\pd^\m\overline{\y})\G^S_{\m\n}\pd^\n \y-m^2\overline{\y}\y .
\ceq
 The introduction of the electromagnetic interaction is standard and
brought about by  the gauge principle leading to the covariant derivative
\aeq
\pd_\m\rightarrow D_\m=\pd_\m+i\charge A_\m,
\ceq
where $\charge$ is the charge of the particle, the resulting gauged Lagrangian being
\aeq
\lag_{\mbox{\footnotesize int}}^S=\lag_{\textup{free}}^S- j_\m A^\m+\charge^2\overline{\y}(\G^S_{\m\n}+\G^S_{\n\m})\y A^\m A^\n .
\ceq
Switching to  momentum space, the electromagnetic current emerges as
\aeq
j^S_{\m}(\mmt',\lambda^\prime;\mmt, \lambda )=\charge \overline{u}(\mmt',\l')(\G^S_{\n\m}p'^\n+\G^S_{\m\n}p^\n)u(\mmt,\l).
\label{spinor_curr}
\ceq
We here are interested in parity conserving interactions and employ the $u({\veg p},\lambda )$, and
$v({\veg p},\lambda )$ Dirac spinors, the amplitudes of the quantum parity states transforming in the $\diracrep$ representation of the
homogeneous Lorentz group (HLG).

Substituting for the explicit form of  $\G^S_{\m\n}$ in (\ref{spinor_curr}), we arrive at
\aeq\label{gordon12}
j^S_{\m}(\mmt',\lambda^\prime;\mmt,\lambda )=\charge \overline{u}(\mmt', \lambda ^\prime)[(p'+p)_\m+i\gs M^{S}_{\m\n}(p'-p)^\n]u(\mmt,\lambda ).
\ceq
Here, in a natural way, the spinless part of the current,  proportional to $(p'+p)^\m$, appears separated from the spin contribution
to the interaction, i.e., from the  $ i \gs M^S_{\m\n}(p'-p)^\n$ term. The expression in (\ref{spinor_curr}) is the counterpart in the NKR formalism
to the Gordon-decomposed Dirac $(D)$ current \cite{Bjorken},
\aeq
j^D_\m(\mmt',\l';\mmt,\l)=\charge\,\overline{u_{D}}(\mmt',\l')[(p'+p)_\m+i 2 M^S_{\m\n}(p'-p)^\n]u_{D}(\mmt,\l),
\label{Dirac12}
\ceq
associated with the phase invariance of the Dirac Lagrangian,
\aeq
\lag_{D}=\overline{\y}(i\g^\m D_\m-m)\y.
\ceq
Notice, that the Gordon decomposed  Dirac current emerges out of the Clifford algebra of the $\gamma$ matrices in combination
with the explicit use of the on-shell  Dirac equation, and can not be obtained directly as a Noether current of a Lagrangian linear
in the momenta. In contrast to this, within the second order formalism discussed here, same expression (\ref{gordon12}) appears directly as a
Noether current due to  phase invariance of the Lagrangian.  However, unlike the Dirac current, where the gyromagnetic ratio is
fixed by the algebra of the $\gamma$ matrices to $g_{(1/2)}=2$,  in (\ref{gordon12}) the value of its counterpart, $\gs$,
in the second  order formalism at first remains unspecified because as usual, Poincar\'e invariance alone correctly
identifies only the mass and the spin of the particle associated with a given representation and is insufficient to fix the values of the
Lagrangian parameters  prior to interactions. The above parameter has been fixed to $\gs =2$ from the requirement on  the asymptotic vanishing  of the Compton scattering cross  section \cite{DelgadoAcosta:2010nx} with energy increase.
%-------------------------------------------------------------------------------------------------------------------------------------
\subsection{The spin 1 current in the $\left( \frac{1}{2}, \frac{1}{2}\right)$ four-vector representation}
%-------------------------------------------------------------------------------------------------------------------------------------
In this case  the equation of motion takes the form \cite{Napsuciale:2007ry}
\aeq
(\G^V_{\a\b\m\n}p^\m p^\n-m^2 g_{\a\b})V^\b=0,
\ceq
where $V^\beta$ denotes the wave function of an ordinary vectorial, i.e.  $J^{P}=1^-$, particle. The   $\G^V_{\a\b\m\n}$ tensor  relevant for parity conserving processes is obtained as,
\aeq
\G^V_{\a\b\m\n}=g_{\a\b}g_{\m\n}-\frac{1}{2}(g_{\a\n}g_{\b\m}+g_{\a\m}g_{\b\n})-i\(\gv-\frac{1}{2}\)[M_{\m\n}^V]_{\a\b},
\label{V_tnsr}
\ceq
where $\gv$ stands for the gyromagnetic ratio of a spin-1 particle as it appears within the covariant projector formalism.
As usual, $M^V_{\m\n}$ are the Lorentz group generators within the  $\procarep$ representation, which are known to be
\aeq
~[M^V_{\m\n}]_{\a\b}=i(g_{\a\m}g_{\b\n}-g_{\a\n}g_{\b\m}).
\label{vector_genr}
\ceq
The associated free Lagrangian reads,
\begin{align}
\lag^V_{\free}=-(\pd^\m V^\a)^{\dagger}\G^V_{\a\b\m\n}\pd^\n V^\b + m^2 V^{\a\dagger}V_\a .
\end{align}
Correspondingly,  the gauged Lagrangian emerges as
\aeq
\lag^V_{\inte}=\lag^V_{\free}- j_\m A^\m-\charge^2 V^{\a\dagger}(\G^V_{\a\b\m\n}+\G^V_{\a\b\n\m})V^\b A^\mu A^\nu.
\ceq
The related electromagnetic  current in momentum space is then obtained as
\aeq
j^V_{\m}(\mmt',\lambda^\prime;\mmt, \lambda)=-\charge {\h^{\a*}}(\mmt', \lambda^\prime)(\G^V_{\a\b\n\m}p'^\n+\G^V_{\a\b\m\n}p^\n)\h^\b(\mmt, \lambda ).
\ceq
Above, $V^\a(x)= \int {\mathrm d}^4p \left(\h^\a({\veg p},\lambda )e^{-i x\cdot p}a _\lambda ({\veg p}) + \h^{\a*}({\veg p},\lambda )
a^\dagger_\lambda ({\veg p}) e^{i x\cdot p}\right)$, and  $\h(\mmt,\lambda )$ are the amplitudes of the spin-1 quantum states that
transform in the $\procarep$ representation of the homogeneous Lorentz group \cite{Bjorken}.
As long as   $p^\a\h_\a(\mmt,\lambda )=0$ holds valid on mass shell,
and using the explicit form of $\G^V_{\alpha\beta\mu \nu}$ in (\ref{V_tnsr}), the expression for the current in momentum space simplifies to
\begin{eqnarray}
j^V_{\m}(\mmt',\lambda^\prime;\mmt, \lambda )&=&-\charge {\h^{\a*}}(\mmt', \lambda^\prime)[(p'+p)_\m g_{\a\b}+i \gv
[M^V_{\m\n}]_{\a\b}(p'-p)^\n-p'_\a g_{\b\m}-p_\b g_{\a\m}]\h^\b(\mmt, \lambda ).\nonumber\\
\end{eqnarray}
In effect, on mass-shell one finds
\begin{eqnarray}\label{gordonv}
j^V_{\m}(\mmt',\lambda^\prime;\mmt,  \lambda )=-\charge {\h^{\a*}}(\mmt',\lambda^\prime )[(p'+p)_\m g_{\a\b}+i\gv [M^V_{\m\n}]_{\a\b}(p'-p)^\n]\h^\b(\mmt ,\lambda ).
\end{eqnarray}
This current again parallels the  Gordon decomposition of the Dirac current,
and, as we shall see below,  coincides with the one used in the standard model  Lagrangian for the $W$ boson.
The current in (\ref{gordonv}) is the counterpart in the NKR formalism to the Proca $(P)$ current,
\aeq
j_\m^P(\mmt',\l';\mmt,\l)=-\charge\,\h^{\a*}(\mmt',\l')[(p'+p)_\m g_{\a\b}-p'_\b g_{\a\m}-p_\a g_{\b\m}]\h^\b(\mmt,\l),
\label{Proca_Gordon}
\ceq
corresponding to the Lagrangian,
\aeq
\lag_P=-\frac{1}{2}[U^{\dagger}]^{\a\b}U_{\a\b}+m^2 [V^{\dagger}]^\a V_\a,\\
\ceq
with $U_{\a\b}=D_\a V_\b-D_\b V_\a$.
Again because of $p^\b \h_\b(\mmt, \l)=0$, this current equivalently rewrites as
\aeq\label{gordonproca}
j_\m^P(\mmt',\l';\mmt,\l)=-\charge\,\h^{*\a}(\mmt',\l')[(p'+p)_\m g_{\a\b}+i[M_{\m\n}^V]_{\a\b}(p'-p)^\n]\h^\b(\mmt,\l).
\ceq
Comparing with Eq.(\ref{gordonv}) we see that Proca's theory comes with a previously  built in gyromagnetic factor of $g_{P}=1$,
a circumstance that will seriously spoil the prediction of the quadrupole moment.

%------------------------------------------------------------------------------------------------------------------------------------
\subsection{The spin 1 current in  the anti-symmetric tensor $(1,0)\oplus(0,1)$   representation}
%------------------------------------------------------------------------------------------------------------------------------------

The   $(1,0)\oplus(0,1)$ single-spin representation has six components and it is well known a fact \cite{Barut:1980aj} that
they can be viewed as the components of a totally anti-symmetric Lorentz tensor of second rank, $F^{\mu\nu}$.
This tensor transforms according to the anti-symmetric part of the direct product of two four-vectors,%

\begin{equation}
\left(  F^{\mu\nu}\right)  ^{\prime}=\frac{1}{2}\left(  \Lambda^\m{}_{\alpha
}\Lambda^{\n}{}_{\beta}-\Lambda_{~\beta}^{\mu}\Lambda
_{~\alpha}^{\nu}\right)  F^{\alpha\beta}\equiv\Lambda_{\quad\alpha\beta
}^{\mu\nu}F^{\alpha\beta},
\label{AStensor}
\end{equation}
where%
\begin{equation}
\Lambda_{~\alpha}^{\mu}=g_{~\alpha}^{\mu}-i(M^V_{\rho\sigma}\frac
{\theta^{\rho\sigma}}{2})_{~\alpha}^{\mu}+{\cal O}(\theta^{2}).
\end{equation}
Here, $\theta^{\rho\sigma}$ stand for the parameters of the Lorentz transformation, with  the
$M^V_{\rho\sigma}$ generators from (\ref{vector_genr}).
Infinitesimally one finds%
\begin{equation}
\Lambda^{\m\n}{}_{\alpha\beta}=\frac{1}{2}\left(  g^\m{}_{\alpha}g^\n{}_{\beta}
-g^\m{}_{\beta}g^\n{}_{\alpha}\right)  -i\left[  \left(
\mathcal{M}^V_{\rho\sigma}\right)^{\m\n}{} _{\alpha\beta}\right]
\frac{\theta^{\rho\sigma}}{2}+O(\theta^{2}),
\end{equation}
which allows to identify the $(1,0)\oplus (0,1)$  generators as%
\begin{equation}
\left(  \mathcal{M}_{\rho\sigma}\right)  _{\quad\alpha\beta}^{\mu\nu}=\frac
{1}{2}\left[  (M^V_{\rho\sigma})_{~\alpha}^{\mu}g_{\nu\beta}+g_{\mu\alpha
}(M^V_{\rho\sigma})_{~\beta}^{\nu}-(M^V_{\rho\sigma})_{~\beta}^{\nu}%
g_{\nu\alpha}-g_{\mu\beta}(M^V_{\rho\sigma})_{~\alpha}^{\mu}\right].
\label{genaten}
\end{equation}
The transformation properties of the components of opposite parities $V_{i}=F^{0i}$ and
$A_{i}=\frac{1}{2}\varepsilon_{ijk}F^{jk}$, spanning this space are now obtained as%
\begin{align}
\mathbf{A}^{\prime} &  =\mathbf{A}+\mathbf{\theta}\times\mathbf{A}%
+\mathbf{\phi}\times\mathbf{V,} \nonumber\\
\mathbf{V}^{\prime} &  =\mathbf{V}-\mathbf{\phi}\times\mathbf{A}%
+\mathbf{\theta}\times\mathbf{V},%
\label{transfav}
\end{align}
where we introduced special notations for the vectorial boost parameters,
$\phi_{i}\equiv\theta_{0i}$, and the pseudo-vectorial rotation parameters, $\theta^{ij}\equiv\varepsilon_{ijk}\theta_{k}$.
The unit operator in this space is
\begin{equation}
\mathbf{1}_{\quad cd}^{ab}=\frac{1}{2}(g^a{}_{c}g^b{}_{d}-g^a{}_{d}g^b{}_{c}),
\end{equation}
while the parity operator reads
\begin{equation}
(\Pi)^{ab}_{\quad cd}=\frac{1}{2}(g_{ac}g_{bd}-g_{ad}g_{bc}), \quad a,b,c,... =0,1,2,3.
\label{parity_asten}
\end{equation}
Notice that in this case we have Lorentz indices associated to the $(1,0)\oplus (0,1)$ representation in
its anti-symmetric tensor form. In order to keep track of the products of operators in this space we use
small Latin letters for the Lorentz indices of the components of the states and operators in the $(1,0)\oplus (0,1)$,
and  hope that such will not lead to confusions.

The explicit connection with the conventional $(1,0)\oplus (0,1)$ states is established upon casting  Eqs. (\ref{transfav})
in matrix form as
\begin{equation}
\left(
\begin{array}
[c]{c}%
i\mathbf{A}^{\prime}\\
\mathbf{V}^{\prime}%
\end{array}
\right)  =\left(
\begin{array}
[c]{cc}%
1-i\mathbf{L\cdot\theta} & \mathbf{L\cdot\phi}\\
\mathbf{L\cdot\phi} & 1-i\mathbf{L\cdot\theta}%
\end{array}
\right)  \left(
\begin{array}
[c]{c}%
i\mathbf{A}\\
\mathbf{V}%
\end{array}
\right),  \label{bvtransf}%
\end{equation}
with $(L_{i})_{jk}=-i\epsilon_{ijk}$, meaning that we can associate to the anti-symmetric tensor the following bi-vector field
\begin{equation}
F^{\mu\nu}\rightarrow \left(
\begin{array}
[c]{c}%
i\mathbf{A}\\
\mathbf{V}%
\end{array}
\right)  ,
\label{bvmap}
\end{equation}
which is Lorentz transformed by the generators
\begin{equation}
\mathbf{J}_{VA}=\left(
\begin{array}
[c]{cc}%
\mathbf{L} & 0\\
0 & \mathbf{L}%
\end{array}
\right)  ,\qquad\mathbf{K}_{VA}=\left(
\begin{array}
[c]{cc}%
0 & i\mathbf{L}\\
i\mathbf{L} & 0
\end{array}
\right)  . \label{genbv}%
\end{equation}
The parity operator within this bi-vector space in the above basis is
\begin{equation}
\Pi_{VA}=\left(
\begin{array}
[c]{cc}%
1 & 0\\
0 & -1
\end{array}
\right)  . \label{pibv}%
\end{equation}

Now we turn to the construction of the Poincar\'e projector. A straightforward calculation using
Eq. (\ref{genaten}) yields %
\begin{align*}
\frac{1}{4}\left(  \mathcal{M}^{\alpha\beta}\mathcal{M}_{\alpha\beta}\right)
_{\quad cd}^{ab} &  =s(s+1)\text{ }\mathbf{1}_{\quad cd}^{ab},\\
\left(  \mathcal{M}_{\mu\alpha}\mathcal{M}^\a{}_{\nu}\right)  _{\quad
cd}^{ab} &  =s(s+1)g_{\mu\nu}\mathbf{1}_{\quad cd}^{ab}-i\left(
\mathcal{M}_{\mu\nu}\right)  _{\quad cd}^{ab},
\end{align*}
with $s=1$. In result, the squared Pauli-Lubanski operator  is obtained as
\begin{equation}
\left(  W^{2}\right)  _{\quad cd}^{ab}=\left(  -\frac{1}{2}\left(
\mathcal{M}^{\alpha\beta}\mathcal{M}_{\alpha\beta}\right)  _{\quad cd}%
^{ab}g^{\mu\nu}+\left(  \mathcal{M}_{\alpha}{}^\mu\mathcal{M}^{\alpha\nu
}\right)  _{\quad cd}^{ab}\right)  p_{\mu}p_{\nu}.
\end{equation}
Correspondingly,  the Poincar\'{e} projector for this representation reads
\begin{equation}
-\frac{W^{2}}{s(s+1)m^{2}}=\left(  g^{\mu\nu}\mathbf{1}+\frac{1}%
{s(s+1)}i\mathcal{M}^{\mu\nu}\right)  \frac{p_{\mu}p_{\nu}}{m^{2}},
\end{equation}
where we now dropped the small Latin letter indices. The anti-symmetric part of the
corresponding tensor remains undetermined  by the Poincar\'{e} projector and as usual
it will be set as the most general anti-symmetric tensor (preserving parity for the purposes
of this work). In so doing,  we find the following  equation of motion,%
\begin{equation}
\left[  \left(  T_{\mu\nu}\right)  _{\quad cd}^{ab}p^{\mu}p^{\nu}-m^{2}%
\mathbf{1}_{\quad cd}^{ab}\right]  F^{cd}=0.
\end{equation}
Here,
\begin{equation}
\left(  T_{\mu\nu}\right)  _{\quad cd}^{ab}=g_{\mu\nu}\mathbf{1}_{\quad
cd}^{ab}-i\gva\left(  \mathcal{M}_{\mu\nu}\right)_{\quad cd}^{ab},
\label{Tmntensor}
\end{equation}
with  $\gva$ being the free parameter, to be associated in the following with the gyromagnetic ratio in
$(1,0)\oplus(0,1)$.
The gauged Lagrangian is then%
\begin{equation}
\mathcal{L}_{\inte}=\left(  D^{\mu}F_{ab}\right)  ^{\dagger}\left(  T_{\mu\nu
}\right)  _{\quad cd}^{ab}D^{\nu}F^{cd}-m^{2}\left(  F^{ab}\right)
^{\dagger}F_{ab},
\end{equation}
which yields the following interactions,
\begin{equation}
\mathcal{L}_{\text{em}}=-ie[\left(  F^{ab}\right)  ^{\dagger}T_{\mu\nu abcd}%
\partial^{\nu}F^{cd}-(\partial^{\nu}F^{ab})^{\dagger}T_{\nu\mu abcd}%
F^{cd}]A^{\mu}+e^{2}\left(  F^{ab}\right)  ^{\dagger}T_{\mu\nu abcd}%
F^{cd}A^{\mu}A^{\nu}.
\end{equation}
The free particle solutions can be  written as usual as %
\begin{equation}
F^{ab}(x)= {\mathcal  F}^{ab} ({\mathbf p},\lambda)e^{-ip\cdot x},
\end{equation}
yielding the electromagnetic current in momentum space as
\begin{equation}
J_{\mu}({\mathbf p},\lambda;{\mathbf p}^{\prime},\lambda^{\prime})=-\charge{\mathcal  F}_{ab}^{\dagger
}({\mathbf p}^{\prime},\lambda^{\prime})\left[  \left(  p^{\prime}+p\right)  _{\mu
}\mathbf{1}_{\quad cd}^{ab}+i\gva \left(  \mathcal{M}_{\mu\nu}\right)  _{\quad
cd}^{ab}(p^{\prime}-p)^{\nu}\right]  {\mathcal  F}^{cd}({\mathbf p},\lambda), \quad \lambda=\pm 1, 0.
\label{currentbv}
\end{equation}

In general, there are six independent degrees of freedom in  $(1,0)\oplus(0,1)$ and we are free to choose the corresponding
basis according to the physics we aim to describe. Concerning electromagnetic interactions,
it is appropriate to work within a basis of well defined parity. Below in subsection {\bf C} of section IV  we shall present
in detail the explicit construction of anti-symmetric tensors describing the negative parity states of interest.
Charge conjugation is especially simple in the tensor basis. Indeed, in position space the field
$F$ satisfies the gauged equation of motion\bigskip%
\begin{equation}
\left[   \left(T_{\mu\nu}\right)  _{\quad cd}^{ab}\left(  \partial^{\mu}+ieA^{\mu}\right)
\left(  \partial^{\nu}+ieA^{\nu}\right)  -m^{2}\mathbf{1}^{ab}_{\quad cd}\right]
F^{cd}=0.\label{eomatf}%
\end{equation}
Taking the complex conjugate of Eq. (\ref{eomatf}) one arrives at%
\begin{equation}
\left[  {T}_{\mu\nu}^{\ast}\left(  \partial^{\mu}-ieA^{\mu}\right)
\left(  \partial^{\nu}-ieA^{\nu}\right)  -m^{2}\right]  F^{\ast}=0,
\end{equation}
where we skipped out the representation indices for simplicity. On the other side,  Eqs. (\ref{genaten}),(\ref{vector_genr}), and
(\ref{Tmntensor}) imply
\begin{equation}
T_{\mu\nu}^{\ast}=T_{\mu\nu},
\end{equation}
and the complex conjugate field satisfies  same gauged equation (\ref{eomatf}) but with an inverse sign of the charge,
\begin{equation}
\left[  T_{\mu\nu}\left(  \partial^{\mu}-ieA^{\mu}\right)  \left(
\partial^{\nu}-ieA^{\nu}\right)  -m^{2}\right]  F^{\ast}=0.
\end{equation}
Hence, the charge conjugated field for this basis is obtained by simple
complex conjugation
\begin{equation}
F^{c}=F^{\ast}.
\label{charge_con_asten}
\end{equation}
It is obvious that the parity operator in (\ref{parity_asten}) commutes with  charge conjugation in (\ref{charge_con_asten}),
thus the charge conjugate states carry equal spatial \ parities.

%------------------------------------------------------------------------------------------------------------------------------------
\subsection{The spin 3/2 current in  the $\left( \frac{1}{2},\frac{1}{2}\right)\otimes
\left[\left( \frac{1}{2},0\right) \oplus \left(0,\frac{1}{2} \right) \right]$ four-vector-spinor representation}
%------------------------------------------------------------------------------------------------------------------------------------

\subsubsection{Covariant projector formalism}

In this case the equation of motion takes the form \cite{Napsuciale:2006wr}
\aeq
(\G_{\a\b\m\n}p^\m p^\n-m^2 g_{\a\b})\y^\b=0.
\label{eomsv}
\ceq
Here,  $\G_{\a\b\m\n}$ contains five independent parity conserving anti-symmetric Lorentz tensors,
weighted by the five free parameters, $c$, $d$, $f$, $g_V$, and $g_S$,
and expresses as \cite{Napsuciale:2006wr}
\begin{align}
\G _{\a\b\m\n}=&-\frac{1}{3} g_{\a\n} g_{\b\m}-\frac{1}{6} i\s_{\a\n} g_{\b\m}-\frac{1}{3}g_{\a\m} g_{\b\n}+\frac{2}{3}g_{\a\b} g_{\m\n}+\frac{i}{3}g_{\m\n}\s_{\a\b}\nonumber\\
&-\frac{i}{6}g_{\b\n}\s_{\a\m}+\frac{i}{6}g_{\a\n}\s_{\b\m}+\frac{i}{6}g_{\a\m}\s_{\b\n}\nonumber\\
&-g_S\frac{i}{2}g_{\a\b}\s_{\m\n}+g_V\left(g_{\a\m}g_{\b\n}-g_{\b\m}g_{\a\n}\right)+id\left(g_{\b\n}\s_{\a\m}-g_{\b\m}\s_{\a\n}\right)\nonumber\\
&+i c\left(g_{\a\m}\s_{\b\n}-g_{\a\n}\s_{\b\m}\right)+if\g^5\e_{\a\b\m\n}. \label{gammageneral}
\end{align}
Contracting the wave equation in (\ref{gammageneral}) first by $p^\a$ and then by $\g^\a$,
amounts to, $-m^2 p_\b \y^\b=0$, and $-m^2\g_\b \y^\b=0$, respectively, meaning that
the  restrictions,
\begin{subequations}\label{restrictions32}
\aec
p^\b \y_\b&=&0,\\
\g^\b \y_\b&=&0,
\cec
\end{subequations}
needed to eliminate the undesired spin $1/2$ sectors are inherent to eq.~(\ref{gammageneral}).
The  associated Lagrangian describing the free negative parity states is
\begin{align}
\lag_{\free}=-(\pd^\m \y^\a)^{\dagger}\G_{\a\b\m\n}\pd^\n \y^\b + m^2\y^{\a\dagger}\y_\a ,
\quad
\psi_\alpha = u_\alpha ({\mathbf p}, \lambda )e^{-ip\cdot x}.
\end{align}
Upon gauging one finds
\aeq
\lag_{\inte}=\lag_{{\free}}-j_\m A^\m-\charge^2\overline{\y}^{\a}(\G_{\a\b\m\n}+\G_{\a\b\n\m})\y^\b A^\mu A^\nu.
\ceq
Then the electromagnetic transition current in momentum space emerges as
\aeq
j_{\m}({\mathbf p}',\lambda^\prime;{\mathbf p}, \lambda )=-\charge \overline{u}^{\a}({\mathbf p}', \lambda^\prime)
(\G_{\a\b\n\m}p'^\n+\G_{\a\b\m\n}p^\n)u^{\b}({\mathbf p},\lambda ), \quad \lambda, \lambda^\prime =\pm \frac{1}{2}, \pm \frac{3}{2},
\ceq
where  $\y^\a=u^\a({\mathbf p},\lambda )e^{-i p\cdot x}$ are the amplitudes of the quantum spin $3/2$
states of negative parity transforming in the $\rsrep$ representation of the HLG.

Some of the five  undetermined parameters in (\ref{eqcoup}) can be fixed by imposing  physical requirements.
In \cite{Napsuciale:2006wr}, the simplest case $f=0$ was studied in detail. Here instead we consider the
most general case of a non-vanishing $f$. We begin with the gauged  equation of motion
\begin{align}\label{eqcoup}
~\left[\G_{\a \b \m \n }D^\m D^\n+m^2 g_{\a \b }\right] \y ^{\b} =
   &{\mathcal D}_{\a\b}\y^{\b}+f(-\g_\a \g_\m \g_\n \g_\b+\g_\a \g_\b g_{\m\n})D^\m D^\n \y^{\b}\nonumber\\
   &+\frac{\charge}{2}(2f+g_S)M_{\m\n}^S g_{\a\b}F^{\m\n}\y^{\b}\nonumber\\
   &+\frac{\charge}{2}\(g_V-c+d-f+\frac{2}{3}\)[M^{V}_{\m\n}]_{\a\b}F^{\m\n}\y^{\b}\nonumber\\
   &+i\charge\(c+f-\frac{1}{6}\)F_{\a \m }\g ^{\m } \g _{\b }\y ^{\b }\nonumber\\
   &-i\charge\(d-f+\frac{1}{6}\)\g _{\a }\g ^{\m } F_{\m \b }\y ^{\b },
\end{align}
with
\begin{equation}
{\mathcal D}_{\alpha\beta} =\left(D^2+m^2\right)g_{\alpha\beta}+\frac{1}{3}\left(\gamma_{\alpha}\gamma ^{\mu } D_{\mu }-4 D_{\alpha }\right)D_{\beta } +\frac{1}{3} \left(D_{\alpha}\gamma ^{\mu }  D_{\mu }-\gamma _{\alpha }D^2\right) \gamma _{\beta }.
\end{equation}

As long as the last two lines of Eq. (\ref{eqcoup}) invoke the undesirable spin-1/2 structure  $(\gamma\cdot \psi)$ into the
 coupling  with the electromagnetic field, we shall eliminate these terms by nullifying their coefficients, yielding
\begin{equation}\label{cfdf}
c=-d=\frac{1}{6}-f.
\end{equation}
In effect,  we are left with
\begin{align}
~\left[\G_{\a \b \m \n }D^\m D^\n+m^2 g_{\a \b }\right] \y ^{\b} =
   &{\mathcal D}_{\a\b}\y^{\b}+f(-\g_\a \g_\m \g_\n \g_\b+\g_\a \g_\b g_{\m\n})D^\m D^\n \y^{\b}\nonumber\\
   &+\frac{\charge}{2}(2f+g_S)M_{\m\n}^S g_{\a\b}F^{\m\n}\y^{\b}\nonumber\\
   &+\frac{\charge}{2}\(g_V+f+\frac{1}{3}\)[M^{V}_{\m\n}]_{\a\b}F^{\m\n}\y^{\b}.
\label{eqcoupfg}
\end{align}
As a next step we aim to identify the spin-3/2 gyromagnetic factor.  In order to ensure proportionality
to $\gsv\veg{S}\cdot\veg{B}$  of the coupling energy of the particle's spin to  the external field,   where $\veg{S}$ stands for
the spin operator within this representation,  we demand the following form of  corresponding interaction Lagrangian,
\begin{equation}\label{EM}
{\cal L}_{\textup{M}}=\gsv\,\frac{\charge}{2}\,\overline{\psi}^\alpha[ M_{\mu\nu}^{3/2}]_{\alpha\beta}\psi^\beta F^{\mu\nu},
\end{equation}
with $\gsv$ denoting the gyromagnetic factor. Given the form of the generators,
\begin{equation}
[M_{\mu\nu}^{3/2}]_{\alpha\beta}=M_{\mu\nu}^{1/2}g_{\alpha\beta}+[M_{\mu\nu}^V]_{\alpha\beta}, \\
\end{equation}
and taking into account that $\g_{\b} \y^{\b}=0$ holds valid to leading order in a perturbative expansion,
one observes that the  second term on the right hand side
of Eq.(\ref{eqcoupfg}) identically vanishes, allowing us to impose the restrictions,
\begin{equation}\label{gpf}
g_{S} =\gsv-2f,\qquad g_V= \gsv-f-\frac{1}{3}.
\end{equation}
With that one is finally  able to write down the tree-level electromagnetic current in terms of a single free parameter,
$\gsv$, as
\begin{equation}
j_\mu(\mmt',\l';\mmt,\l)=-\charge\,\overline{u}^{\alpha}(\mmt',\l')\[ g_{\alpha\beta}(p'+p)_\mu+i \gsv[M^{3/2}_{\mu\nu}]_{\alpha\beta}(p'-p)^\nu\] u^\beta(\mmt,\l).
\label{gordonNKR}
\end{equation}
The conclusion is that in the NKR formalism,  the spin-3/2 current in the four-vector spinor associated with local phase invariance
naturally decomposes into a motion (convection) part and a spin-magnetization part, which, as we will see below,
in reality provides contributions to all the allowed higher multipoles of the particle. As a reminder,
in \cite{Napsuciale:2006wr}, it was shown that the causal
propagation of spin $3/2$ waves in an electromagnetic background requires $\gsv=2$, a conclusion also independently
verified  by the unitarity
of the differential cross section in the forward direction for Compton scattering \cite{DelgadoAcosta:2009ic}.

The expression in (\ref{gordonNKR}) is the counterpart in the NKR formalism of the Gordon decomposition of the Rarita-Schwinger current
\begin{eqnarray}
j_\m^{RS}(\mmt',\lambda';\mmt,\lambda)&=&-2m \charge\,\overline{u}^{\alpha}
(\mmt',\lambda') \gamma_\mu  u_\a(\mmt,\lambda),\nonumber\\
&=&-\charge\,\overline{u}^{\alpha}(\mmt',\lambda')
 [(p'+p)^\m+ig_{(1/2)} M^{1/2}_{\m\n}(p'-p)^\n]  u_\a(\mmt,\lambda).
\label{RS_gordon}
\end{eqnarray}
Notice  that the principal difference between Eqs.~(\ref{gordonNKR}) and (\ref{RS_gordon}) is due to
the appearance of the purely Dirac spin-magnetization tensor, $g_{(1/2)} M^{1/2}_{\m\n}$
in (\ref{RS_gordon}) in place of the genuine four-vector--spinor spin-magnetization tensor,
$\gsv[M^{3/2}_{\mu\nu}]_{\alpha\beta}$, in (\ref{gordonNKR}). In order to illuminate the  origin
of this crucial difference we take in the next section a closer look on the eigenvalue problem of the squared Pauli-Lubanski vector
in $\psi_\mu$ and its relationship  to the Rarita-Schwinger formalism for spin $3/2$.

\subsubsection{Shortcomings of the Rarita-Schwinger framework from the perspective of the covariant projector}

We begin by recalling that in the $(1/2,1/2)\otimes [(1/2,0)\oplus(0,1/2)]$ direct product space,
the principal ingredient of the respective covariant projector, the Pauli-Lubanski vector,
${\mathcal W}_\mu$, is obtained
as the direct sum of the Pauli-Lubanski vectors, $W_\mu$,  and $w_\mu$, in the respective  $(1/2,1/2)$- and
Dirac-building blocks according to,
\begin{eqnarray}
\left[\left[{\mathcal W}_\mu\right]_\alpha{}^\beta\right]_{AB} &=&
\left[ w_\mu\right]_{AB} g_\a{}^\beta  + \left[W_\mu\right]_\alpha{}^\beta\delta_{AB},\nonumber\\
w_\mu=\frac{1}{2}\gamma_5(p_\mu -\gamma_\mu  p\!\!/), &\quad&
\left[ W_\mu\right]_{\alpha}{}^{\beta}=i\epsilon_{\mu\alpha}{}^\beta{}_{\sigma}p^\sigma.
\label{PaLu_VS}
\end{eqnarray}
The squared Pauli-Lubanski vector in  $\psi_\mu$ is then calculated as
\begin{subequations}\label{old29}
\aec
\left[ {\mathcal W}^2\right]_\alpha{} ^\beta &=&
w^2g_\alpha{}^\beta +\left[W^2\right]_\alpha{}^\beta  + 2(W^\mu)_\alpha{}^\beta w_\mu,\label{Pl1}
\\
w^2&=&-\frac{1}{4}\sigma_{\lambda\mu} \sigma^\lambda{}_\nu p^\mu p^\nu, \label{Pl2}
\\
\left[W^2 \right]_\alpha{}^\beta&=&
-2(g_{\alpha}{}^{\beta}g_{\mu\nu}  -g_{\alpha\nu}g^{\b}{}_{\mu})p^\mu p^\nu,\\
2(W^\mu)_\alpha{}^\beta w_\mu &=& i\epsilon^\mu{}_\alpha{}^\beta{}_\sigma p^\sigma\gamma_5(p_\mu-\gamma_\mu p\!\!/)
=-i\epsilon^\mu{}_\alpha{}^\beta{}_{\s} p^\sigma \gamma_5\gamma_\mu p\!\!/,
\label{PL3}
\cec
\end{subequations}
where for the sake of simplicity we suppressed the spinorial indices.
Then the wave equation takes the form in Eq. (\ref{eomsv}) with
\begin{equation}
\Gamma_{\alpha\beta\mu\nu}=\frac{2}{3}\left(  g_{\alpha\beta}g_{\mu\nu}-g_{\alpha\nu}g_{\beta\mu}\right)
+\frac{1}{6}(\epsilon^\l{}_{\alpha\beta\mu}\gamma^{5}\sigma_{\lambda\nu}
+\epsilon^\l{}_{\alpha\beta\nu}\gamma^{5}\sigma_{\lambda\mu})
+\frac{1}{12}\sigma_{\lambda\mu}\sigma^\l{}_{\nu}g_{\alpha\beta}-\frac{1}{4}g_{\mu\nu}g_{\alpha\beta}\,.
\label{old2}
\end{equation}
The first and third terms in (\ref{old2}) take their origins in  turn from
the contributions to the projector of $W^2$, and  $w^2$  according to (\ref{old29}), while
the second term arises due to the contribution of the interference term, $2W\cdot w$,
in (\ref{PL3}).
\noindent
On mass shell,   (\ref{old2}) leads to
\begin{eqnarray}
(i\epsilon_{\alpha\beta\mu\sigma}\gamma^5\gamma^\mu p^\sigma p\!\!/ -m^2g_{\alpha\beta}+2p_\beta p_\alpha)\psi^{\(m,\frac{3}{2}\)}\, ^\beta
=0.
\label{RS_inc}
\end{eqnarray}
The latter  equation bears strong resemblance to a version of the Rarita-Schwinger equation of frequent use \cite{Lurie},
\begin{equation}
(i\epsilon_{\alpha\beta\mu\sigma}\gamma^5\gamma^\mu p^\sigma -m g_{\alpha\beta}+ \gamma_\a\gamma_\b)\psi^{\(m,\frac{3}{2}\)}\, ^\beta
=0,
\label{Lurie}
\end{equation}
as visible upon substituting $\gamma\cdot \psi=0$  in (\ref{Lurie}) by the more fundamental
$p\cdot \psi=0$ , and $p\!\!/\psi^{(m,3/2)}=m\psi^{(m,3/2)}$.
The above considerations show that the Rarita-Schwinger framework solely captures
the piece of the covariant projector given by the interference term in
(\ref{old2}) while ignoring  the rest. The omission of the $W^2$ and $w^2$ contributions
to (\ref{old2}) and the subsequent linearization of the covariant projector by the Rarita-Schwinger framework,
seriously prejudices the coupling of spin-3/2 to the electromagnetic field and is at the root of the
inconsistencies of the interacting Rarita-Schwinger theory.

%---------------------------------------------------------------------------------------------------------------
\subsection{Spin $3/2$ current in the single-spin  $\tmrep$ representation}
%---------------------------------------------------------------------------------------------------------------

Even though the most interesting cases of spin $3/2$ particles belong to the $\rsrep$ representation, we include
for completeness the multipole moments of the $\tmrep$ representation as well. To obtain the current for this representation
we need to work out the corresponding Poincar\'e projector whose calculation can be done for an arbitrary half-integer
spin  $j$ residing in the
$(j,0)\oplus (0,j)$ representation.

In choosing a rotational  $\{|j,\lambda \rangle\}$ basis within  anyone of the  $(j,0)$, and $(0,j)$ subspaces, the
generators for the $(j,0)\oplus(0,j)$ representation can be written as%
\begin{equation}
\mathbf{J}=\left(
\begin{array}
[c]{cc}%
\mathbf{J}^{(j)} & 0\\
0 & \mathbf{J}^{(j)}%
\end{array}
\right)  ,\qquad\mathbf{K}=\left(
\begin{array}
[c]{cc}%
i\mathbf{J}^{(j)} & 0\\
0 & -i\mathbf{J}^{(j)}%
\end{array}
\right).  \label{genchi}%
\end{equation}
Here, $\mathbf{J}^{(j)}$ stand for the conventional $(2j+1)\times(2j+1)$ rotation matrices for spin-$j$ in the $\{|j,\l\rangle\}$ basis. The $(j,0)$,  and $(0,j)$ states are in turn associated with right- and left-
chiralities. In
momentum space one encounters
\aeq
\psi ({\mathbf p},\l )=\left(
\begin{array}[c]{c}
\phi_{R} ({\mathbf p},\l ) \\
\phi_{L} ({\mathbf p},\l )
\end{array}
\right), \quad \lambda=\pm \frac{1}{2}, ...,\pm j,
\label{chistates}
\ceq
a spinor which transforms according to
\begin{equation}
\Lambda=\left(
\begin{array}
[c]{cc}%
\Lambda_{R}&0\\
0 & \Lambda_{L}%
\end{array}
\right),
\end{equation}
where  the $\Lambda_{R}$--, and $\Lambda_L$ matrices  transform the respective
right--($(j,0)$), and left-handed ($(0,j)$) subspaces,  whose generators were given
in the diagonals of the matrices in  Eq.(\ref{genchi}). The generators of the combined  $(j,0)\oplus(0,j)$ representation satisfy
\aeq
\mathbf{K}^{2}=-\mathbf{J}^{2}.
\ceq
The conventional Lorentz  generators, $M^{\mu\nu}$,
are now identified  as  $M^{0i}=K_{i}$, $M^{ij}=\varepsilon_{ijk}J_{k}$
and it is straightforward to prove that they satisfy the standard Lorentz algebra%
\aeq
\lbrack M^{\mu\nu},M^{\alpha\beta}]=-i\left(  g^{\mu\alpha}M^{\nu\beta}%
-g^{\mu\beta}M^{\nu\alpha}-g^{\nu\alpha}M^{\mu\beta}+g^{\nu\beta}M^{\mu\alpha
}\right)  .
\ceq
Squaring the Pauli-Lubanski vector,
\aeq
W_{\alpha}=\frac{1}{2}\varepsilon_{\alpha\rho\sigma\mu}M^{\rho\sigma}p^{\mu},
\ceq
amounts to
\aeq
W^{2}=\left(  -\frac{1}{2}M^{\alpha\beta}M_{\alpha\beta}g^{\mu\nu}
+M_\a{}^{\mu}M^{\alpha\nu}\right)  p_{\mu}p_{\nu}.
\ceq
A straightforward calculation with the generators in Eq. (\ref{genchi}) yields for any $j$, %
\begin{align*}
\frac{1}{2}M^{\alpha\beta}M_{\alpha\beta} &  =\left(  \mathbf{J}%
^{2}-\mathbf{K}^{2}\right)  =2\mathbf{J}^{2}=2j(j+1),\\
M_{\alpha}{}^{\mu}M^{\alpha\nu} &  =j(j+1)g^{\mu\nu}-iM^{\mu\nu}.%
\end{align*}
Correspondingly,  the Poincar\'{e} projector emerges as
\aeq
{\cal P}^{(m,s)}=\frac{p^{2}}{m^{2}}\left[  -\frac{W^{2}}{j(j+1)p^{2}}\right]
=-\frac{W^{2}}{j(j+1)m^{2}}.
\ceq
As long as the anti-symmetric part of this operator remains unspecified by Poincar\'e invariance,
we as usual shall take advantage of this freedom and extend  the Lagrangian by
the most general covariant anti-symmetric tensor. In so doing,   the
following equation of motion in the representation under consideration is found,
\aeq
\left(  T_{\mu\nu}p^{\mu}p^{\nu}-m^{2}\right)  \psi=0,
\ceq
with
\aeq
T_{\mu\nu}=g_{\mu\nu}-igM_{\mu\nu}.
\ceq
This tensor depends on one sole free parameter, $g$,
provided, we have restricted the formalism to parity conserving processes.

The Lagrangian for a particle in the $(j,0)\oplus(0,j)$ representation
interacting with an electromagnetic field is%

\begin{equation}
\mathcal{L}_{\inte}=\overline{D^{\mu}\psi}T_{\mu\nu}D^{\nu}\psi-m^{2}\overline{\psi
}\psi,
\label{Gl_int}
\end{equation}
where $D^{\mu}=\partial^{\mu}+ieA^{\mu}$, and $e$ is the charge of the
particle. Here, one defines the adjoint spinor  as%
\aeq
\overline{\psi}=\psi^{\dagger}\Pi,
\ceq
with $\Pi$ being the parity operator, which in the chiral basis coincides with  the off-diagonal
$6\times 6$ unit matrix,
\begin{equation}
\Pi=\left(
\begin{array}
[c]{cc}%
0 & 1\\
1 & 0
\end{array}
\right)  .\label{pichiconv}%
\end{equation}
Explicitly, the  electromagnetic interactions described by ${\mathcal L}_{\inte}$ in (\ref{Gl_int})  are,
\begin{equation}
\mathcal{L}_{\inte}=-ie[\overline{\psi}T_{\mu\nu}\partial^{\nu}\psi
-(\partial^{\nu}\overline{\psi})^{\dagger}T_{\nu\mu}\psi]A^{\mu}%
-e^{2}\overline{\psi}T_{\mu\nu}\psi A^{\mu}A^{\nu}.
\end{equation}
Using the momentum space functions, $w({\mathbf p},\lambda)e^{-ip\cdot x}$,  with $\lambda=\pm \frac{1}{2}, ...,\pm j$,
yields  the following electromagnetic current
\aeq
j^{(j)}_\mu({\mathbf p}',\l';{\mathbf p},\l)=\charge\,\overline{w}({\mathbf p}',\l')\left[ (p'+p)_\m+ig^{(j)} M^{(j)}_{\m\n}(p'-p)^\n\right]
w({\mathbf p},\lambda).
\label{jcurrent}
\ceq
Here, we attached the representation index $j$ to  the generators, the free parameter, and the current.
The states $w({\mathbf p},\lambda)$ of well defined parity are obtained from
the chiral states in Eq. (\ref{chistates}) by multiplication by the matrix
\aeq
M=\frac{1}{\sqrt{2}}
\left(
\begin{array}
[c]{cc}%
1 & 1\\
1 & -1%
\end{array}
\right).
\ceq
Consequently, the Lorentz transformations of the parity basis express in terms of the Lorentz transformations of the chiral states as
\begin{equation}
\Lambda=\frac{1}{2}\left(
\begin{array}
[c]{cc}%
\Lambda_{R}+\Lambda_{L} & \Lambda_{R}-\Lambda_{L}\\
\Lambda_{R}-\Lambda_{L} & \Lambda_{R}+\Lambda_{L}%
\end{array}
\right)  .
\end{equation}
Finally, before closing this section and for the needs  of what follows
we wish  to bring the boost operator in the $(3/2,0)\oplus (0,3/2)$ in this basis,
\begin{equation}
B(\mathbf{p})=\left(
\begin{array}
[c]{cc}%
\cosh\left(  \mathbf{J}\cdot\mathbf{n}\varphi\right)  & \sinh\left(
\mathbf{J}\cdot\mathbf{n}\varphi\right) \\
\sinh\left(  \mathbf{J}\cdot\mathbf{n}\varphi\right)  & \cosh\left(
\mathbf{J}\cdot\mathbf{n}\varphi\right)
\end{array}
\right).
\label{boost32}
\end{equation}

%-------------------------------------------------------------------------------------------------
%-------------------------------------------------------------------------------------------------
\section{Multipole Expansions }
%-------------------------------------------------------------------------------------------------
%-------------------------------------------------------------------------------------------------
%--------------------------------------------------------------------------------------------------
The electromagnetic moments of a particle are defined by means of a multipole expansion of a
corresponding current density. The current densities used here are obtained in transforming the
electromagnetic currents from above  to the Breit (B) frame:
\aeq
J^B_\m(\veg{q},s,\l)=\frac{1}{\o}j_\m^{(s)}(\mmt',\l;\mmt,\l),\qquad p'
=({\o}/{2},{\veg{q}}/{2}),\quad p=({\o}/{2},-{\veg{q}}/{2}),
\ceq
with $\omega=\sqrt{4m^2 +\veg{q}^2}$.
Expressions for the  Cartesian electromagnetic moments for a particle of spin $s$ and polarization $\l$
can be found in  \cite{Kleefeld:2000nv} and read,
\aec
Q_{E}^l(\veg{q}, s,\l)&=&\left.b^{l0}(-i\pd_{\veg{q}})
\varrho_E(\veg{q}, s,\l)\right\vert_{\veg{q}=0},
\label{qeformulas}\\ Q_{M}^l(\veg{q}, s,\l)
&=&\frac{1}{l+1}\left.b^{l0}(-i\pd_{\veg{q}})
{\varrho_M}(\veg{q}, s,\l)\right\vert_{\veg{q}=0},
\label{qmformulas}
\cec
where the electric density $\varrho_E(\veg{q}, s,\l)$, and the  magnetic density
$\varrho_M(\veg{q}, s,\l)$ are,
\aeq\label{bdensities}
\varrho_E(\veg{q}, s,\l)=j_B^0(\veg{q}, s,\l),\qquad \varrho_M(\veg{q}, s,\l)
=\pd_{\veg{q}}\cdot[\veg{j}_B(\veg{q},s,\l)\times\veg{q}].
\ceq
The $b^{l0}(-i\pd_{\veg{q}})$ operators  can be obtained from their definitions in position space
in terms of the spherical harmonics,
\aeq
b^{l0}(\veg{r})=l !\sqrt{{4\p}/(2l+1)}r^l Y_{l0}(\Omega),
\ceq
upon Fourier transformations toward momentum space,
\aec
b^{00}(-i\pd_{\veg{q}})&=&1,\\
b^{10}(-i\pd_{\veg{q}})&=&-i\frac{\pd}{\pd q_z},\\
b^{20}(-i\pd_{\veg{q}})&=&
\frac{\pd^2}{\pd q_x^2}+\frac{\pd^2}{q_y^2}-2\frac{\pd^2}{\pd q_z^2},\\
b^{30}(-i\pd_{\veg{q}})&=&
-3\,i\,\frac{\pd}{\pd q_z}\(3\frac{\pd^2}{\pd q_x^2}
+3\frac{\pd^2}{\pd q_y^2}-2\frac{\pd^2}{\pd q_z^2}\),\quad \text{etc}.
\label{bl0}
\cec

%%%%%%%%%%%%%%%%%%%%%%%%%%%%%%%%%%%%%%%%%%%%%%%%%%%%%%%%%%%%%%%%%%%%%%%%%%%%%%%%%%%%%%%%%%%%%%%%%%%%%%%%%%%%%%%%%%%%%%%%%%%%%%%%%%%%%%%%%%%%
\section{Electromagnetic multipoles  from the second order formalism}
%%%%%%%%%%%%%%%%%%%%%%%%%%%%%%%%%%%%%%%%%%%%%%%%%%%%%%%%%%%%%%%%%%%%%%%%%%%%%%%%%%%%%%%%%%%%%%%%%%%%%%%%%%%%%%%%%%%%%%%%%%%%%%%%%%%%%%%%%%%%
In this section we perform side by side  calculations of
the electromagnetic multipole moments of elementary particles with spins 1/2, 1, and 3/2 within the
Poincar\'e covariant second order (NKR) formalism and the respective Dirac-, Proca-, and Rarita-Schwinger frameworks.

%----------------------------------------------------------------------------------------------------------------------------------------------------

\subsection{Spin $1/2$ in  $\diracrep$ }
We first examine the multipole decompositions of the currents in
Eqs.~(\ref{gordon12}), and (\ref{Dirac12}) under employment of the  well known spinors,
\begin{eqnarray}
u\left(\veg{p},+\frac{1}{2}\right)= N\left(
\begin{array}{c}
m+p_0 \\
0 \\
p_z \\
p_x+i p_y%
\end{array}
\right),&&u\left(\veg{p},-\frac{1}{2}\right)= N
\left(
\begin{array}{c}
0 \\
m+p_0 \\
p_x-i p_y \\
-p_z%
\end{array}
\right),
\label{base12}
\end{eqnarray}
with $N=[2m(m+p_0)]^{-1/2}$. The Breit frame densities  (\ref{bdensities}) corresponding to the current in
Eq.~(\ref{gordon12})
are found as
\begin{subequations}\label{Ddensities122}
\begin{eqnarray}
\varrho_E^S(\veg{q},1/2,\pm 1/2)&=&\frac{\charge(8m^2-(\gs-2)\veg{q}^2)}{4m \o},\\
\varrho_M^S(\veg{q},1/2,\pm 1/2)&=&\pm \frac{i \charge \gs q_z}{\o},
\end{eqnarray}
\end{subequations}
The associated multipoles are then,
\begin{subequations}\label{DMoments12}
\begin{align}
\(Q^0_E\)^{NKR}&=\charge,\\
\(Q^1_M\)^{NKR}&=\frac{\charge \gs }{2m}\langle S_z\rangle,
\end{align}
\end{subequations}
where we have used the following compact notations,
\begin{subequations}\label{simpnot}
\aec
Q_E^0&\equiv& Q_E^0(s,\l),\\
\langle\,{\mathcal O}\,\rangle &\equiv& \langle\,s,\l \vert{\mathcal O}\vert s,\l \,\rangle,
\cec
\end{subequations}
with $s=\frac{1}{2},\, \l=\pm \frac{1}{2}$. In Eqs. (\ref{DMoments12}),  $Q^0_E$ and $Q^1_M$ denote  in turn the electric
monopole  and  magnetic dipole moments, all other moments  vanish.
The above expressions reproduce the electric monopole and the magnetic dipole moments of the Dirac electron for $\gs=2$.
Within the NKR method, the $g_S$ value has been fixed to $\gs=2$ in \cite{DelgadoAcosta:2010nx} from the requirement of
reproducing the correct asymptotic behavior of the Compton scattering cross sections with energy increase.
Therefore,  the results on the multipole moments for  a
fundamental spin 1/2 particle predicted by the covariant-projector formalism
are the same as those following from the Dirac Lagrangian.
However, this is not to be so for the higher spins $s=1$,
and $s=3/2$, where one detects problems in the Proca and Rarita-Schwinger approaches, which need special attention.

%--------------------------------------------------------------------------------------------------------------------------------------
\subsection{Spin $1$  in  $\procarep$}
%--------------------------------------------------------------------------------------------------------------------------------------
Next we  turn to the vector case. We shall be calculating the multipole expansions of the currents in Eqs.~(\ref{gordonv}),
and (\ref{Proca_Gordon}). The basis vectors, $\eta ({\mathbf p},\lambda)$, with $\lambda =\pm 1,0$,  used by us  are,
\begin{subequations}\label{base1}
\begin{eqnarray}
\eta(\veg{p},1)&=&\frac{\mathcal{N}}{\sqrt{2}} \left(
\begin{array}{c}
-\left(m+p_0\right) \left(p_x+i p_y\right) \\
-m^2-p_0 m-p_x^2-i p_x p_y \\
-i \left(p_y^2-i p_x p_y+m \left(m+p_0\right)\right) \\
-\left(p_x+i p_y\right) p_z%
\end{array}
\right),\\
\eta(\veg{p}, 0)&=&\mathcal{N} \left(
\begin{array}{c}
\left(m+p_0\right) p_z \\
p_x p_z \\
p_y p_z \\
p_z^2+m \left(m+p_0\right)%
\end{array}%
\right), \\
\eta(\veg{p}, -1)&=&\frac{\mathcal{N}}{\sqrt{2}} \left(
\begin{array}{c}
\left(m+p_0\right) \left(p_x-i p_y\right) \\
m^2+p_0 m+p_x^2-i p_x p_y \\
-i \left(p_y^2+i p_x p_y+m \left(m+p_0\right)\right) \\
\left(p_x-i p_y\right) p_z%
\end{array}
\right),
\end{eqnarray}
\end{subequations}
with $\mathcal{N}=[m(m+p_0)]^{-1}$.
The transverse densities are now calculated as,
\begin{subequations}\label{NKR_MM1}
\aec
\varrho_E^V(\veg{q},1,\pm 1)&=&\frac{\charge(4m^2+q_x^2+q_y^2)(8m^2+4 \o m+\veg{q}^2)}{4m^2(2m+\o)^2}\nonumber\\
&&-\frac{\charge \gv(q_x^2+q_y^2)(4m^2+2\o m+\veg{q}^2)}{4m^2(2m+\o)^2}\\
&=& \frac{\charge[4m^2-(\gv-1)(q_x^2+q_y^2)]}{4m^2},\\
\varrho_M^V(\veg{q},1,\pm 1)&=&\pm \frac{i \gv \charge q_z(4m^2+2\o m+\veg{q}^2)}{m\o(2m+\o)}=\pm \frac{i \gv \charge q_z}{m}.
\cec
\end{subequations}
For the longitudinal densities ($\l=0$) we find,
\begin{subequations}
\aec
\varrho_E^V(\veg{q},1,0)&=&\frac{\charge(2m^2+q_z^2)(8m^2+4 \o m+\veg{q}^2)}{2m^2(2m+\o)^2}-\frac{\charge \gv q_z^2(4m^2+2\o m+\veg{q}^2)}{2m^2\o(2m+\o)}\\
&=&\frac{\charge[2m^2-(\gv-1)q_z^2]}{2m^2},\\
\varrho_M^V(\veg{q},1,0)&=&0.
\cec
\end{subequations}
Correspondingly, in terms of the expectation values of the $\veg{S}^2$ and $S_z^2$ operators, one encounters,
\begin{subequations}\label{NKR_hfhf}
\begin{align}
\(Q_E^0\)^{V}&=\charge,\\
\(Q_M^{1}\)^{V}&=\frac{\charge \gv}{2m}\langle S_z\rangle,\\
\(Q_E^{2}\)^{V}&=\frac{\charge(1-\gv) }{m^2}\langle 3S_z^2-\veg{S}^2\rangle,
\end{align}
\end{subequations}
where we have adopted the notation (\ref{simpnot}) with $s=1$ in this case. The $\gv$ value has been fixed
in ref.~\cite{Napsuciale:2007ry} to $\gv=2$ from the requirement on
unitarity of the Compton scattering amplitudes in the ultraviolet.
All multipoles are non-vanishing and interrelated by the $\gv$ value, as it should be,
due to the impossibility to separate the electric and magnetic  fields covariantly.
Thus the NKR method predicts a negative electric quadrupole moment of the $W$ boson in agreement with the
Standard Model (see for example \cite{Lee:1981mf, Honzawa:1991qn, Holstein:2006wi}) and in line with the empirical observations
\cite{Abreu:2001rpa}. In contrast to this,  the Proca Lagrangian predicts a vanishing electric quadrupole, as visible upon substituting
$\gv=1$ in (\ref{NKR_MM1}). This  shortcoming of Proca's framework is removed by the standard model
on the cost of introducing a non-Abelian current \cite{Holstein:2006wi},
\begin{equation}
J_\mu^{\textit{NA}}(\veg{p}^\prime,\lambda^\prime;\veg{p},\lambda)=-i\charge\,\eta^\ast\, ^\alpha (\veg{p}^\prime, \lambda^\prime)
\left[ M^V_{\mu\nu}\right]_{\alpha\beta}(p^\prime-p)^\n\eta^\beta(\veg{p},\lambda),
\end{equation}
which leads to a gyromagnetic $W$-boson ratio of $g_W=2$. The r\'ole of this current is same as in the NKR method, namely, removing the incompleteness of
Proca's Lagrangian brought about by the omission of the most general anti-symmetric  tensor allowed by Poincar\'e
invariance that respects the exclusion of the redundant spin-0 (time-like) component of the four-vector.
In effect, tree-level  Compton scattering within the NKR method results are the same as within the Standard Model, though the former is valid for
any spin-1 in $(1/2,1/2)$, no matter Abelian, or non-Abelian (c.f. ~\cite{Napsuciale:2007ry} for details).

%--------------------------------------------------------------------------------------------------------------------------------------
\subsection{Spin $1$  in  $(1,0)\oplus (0,1)$ }
%--------------------------------------------------------------------------------------------------------------------------------------
In the current section we explore the representation dependence of the multipole moments of a vector particle.
For this purpose we study the electromagnetic properties of a fundamental single-spin 1 residing in the
$(1,0)\oplus (0,1)$ representation space.
In the following, this representation  will be termed to  either as  anti-symmetric second-rank tensor, or, equivalently,
as  bi-vector. In order to calculate the multipole decomposition of  the current given in (\ref{currentbv}) above,
we need to explicitly construct the ${\mathcal  F}^{\mu\nu}$  tensor components, equivalently, the ${\mathbf A}$
and ${\mathbf V}$ vectors in (\ref{transfav}), describing spin 1 particles. This can
be easily done with the aid of the boost generators in Eq.( \ref{genbv}). In so doing, one finds the boost operator as
\begin{equation}
B(\mathbf{p})=\left(
\begin{array}
[c]{cc}%
\cosh\left(  \mathbf{L}\cdot\mathbf{n}\varphi\right)  & \sinh\left(
\mathbf{L}\cdot\mathbf{n}\varphi\right) \\
\sinh\left(  \mathbf{L}\cdot\mathbf{n}\varphi\right)  & \cosh\left(
\mathbf{L}\cdot\mathbf{n}\varphi\right)
\end{array}
\right).
\label{boostbv}
\end{equation}
Using $(\mathbf{L}\cdot\mathbf{n})^{3}=\mathbf{L}\cdot\mathbf{n}$ we find%
\begin{align}
\cosh\left(  \mathbf{L}\cdot\mathbf{n}\varphi\right)   &  =1 +\left(  \mathbf{L}%
\cdot\mathbf{n}\right)  ^{2}\left(  \cosh\varphi-1\right)  ,\\
\sinh\left(\mathbf{L}\cdot\mathbf{n}\varphi\right)&  =\mathbf{L}%
\cdot\mathbf{n}\left(  \sinh\varphi\right)  .
\end{align}
A straightforward calculation yields
\begin{equation}
\left(  \mathbf{L}\cdot\mathbf{n}\right)  _{ij}   =-i\epsilon
_{ijm}n_{m} \qquad
\left(  \mathbf{L}\cdot\mathbf{n}\right)  _{ij}^{2}
=\left(  \mathbf{n}^{2}\delta_{ij}-n_{i}n_{j}\right) ,
\end{equation}
which allows us to explicitly construct the boost for the bi-vector state and therefore to produce explicit
expressions for the corresponding $\mathbf{A}, \mathbf{V}$ components. The anti-symmetric tensor
describing negative parity spin $1$ particles  obtained in this fashion has the following components
\begin{align}
\left( i \mathbf{A}_{(\z)}(\mathbf{p})\right)  _{i}  &  =-i\epsilon_{i\z m}%
\frac{p_{m}}{m},\label{VAstates}\\
\left(  \mathbf{V}_{(\z)}(\mathbf{p})\right)  _{i}  &  =\frac{p_{0}}{m}\delta_{i\z}%
-\frac{p_{i}p_{\z}}{m\left(  p_{0}+m\right)  },\nonumber
\end{align}
with $\z=1,2,3$ labeling the three independent bi-vectors.
The set of three ${\mathcal  F}^{\mu\nu}_{(\z)}$ tensors calculated in this way is equivalent to  the following construct (see refs.~\cite{Acosta:2004pk} and
references therein for more details), in terms of the $(1/2,1/2)$ spinors
$\eta (\mmt, \lambda )$ in Eqs.~(\ref{base1}),
\aeq\label{F-Ta}
~{\mathcal F}^{\m\n}_{(\z)}(\mmt)=e_0^{\m}(\mmt)e_{(\z)}^{\nu}(\mmt)-e_{(\z)}^{\m}(\veg{p})e_0^{\nu}(\mmt),
\ceq
with
\begin{subequations}
\label{mom_stts}
\begin{eqnarray}
e_0^\a(\mmt)&=&\frac{1}{m}p^\a,\\
e_{(x)}^\a(\veg{p})&=&\frac{1}{\sqrt{2}}(\h^\a(\mmt,-1)-\h^\a(\mmt,+1)),\\
e_{(y)}^\a(\veg{p})&=&\frac{i}{\sqrt{2}}(\h^\a(\mmt,-1)+\h^\a(\mmt,+1)),\\
e_{(z)}^\a(\veg{p})&=&\h^\a(\mmt,0).
\end{eqnarray}
\end{subequations}
Notice, that at rest,  $\mathbf{V}_{(\z )}$ reduce to the three Cartesian unit vectors ,
$\hat{\mathbf{e}}_{(\z)}$, while $\mathbf{A}_{(\z )}$  vanish.
We will not discuss here in detail the positive parity solutions, but they can be obtained  as the dual tensors
to the ones in Eq.(\ref{F-Ten}).

It has to be noticed that a set of six  bi-vectors spanning the $(1,0)\oplus (0,1)$ representation space has
already been constructed earlier  in \cite{Ahluwalia:1999ny} though on the grounds of a different representation of the
$J_i$ generators. The main purpose in \cite{Ahluwalia:1999ny} has been the construction of Feynman propagators from the explicit
construction of the states and without reference to a Lagrangian formalism. The problem  of the identification of
the degrees of freedom of the corresponding anti-symmetric tensor has not been addressed there.
We here instead are interested in  the electromagnetic couplings of general spin 1 particles, including composite ones, and
entirely focus on the $ (1,0)\oplus (0,1)\sim {\mathcal F}^{\mu\nu }_{(\z)}$ map.  We  construct an
electromagnetic current that follows from the Lagrangian underlying  the covariant projector
wave equation for  ${\mathcal F}^{\mu\nu }_{(\z)}$.
An additional motivation to work with this field is the straightforward generalization of the dimensional regularization
method in the calculation of quantum corrections (work in progress).

In the calculation of the multipole moments we will need the explicit form of the states, ${\mathbf A}({\mathbf p}, \lambda)/{\mathbf V}({\mathbf p},\lambda)$,
of well defined parity and angular momentum, which  relate to  the Cartesian states
${\mathbf A}_{(\zeta)}({\mathbf p})/{\mathbf V}_{(\zeta)}({\mathbf p })$,  from (\ref{VAstates}) by combinations of the type given in (\ref{mom_stts}).
Then the three corresponding anti-symmetric tensors are,
\begin{equation}
{\mathcal F}^{\mu\nu}(\mmt,\l)=\left(
\begin{array}{cccr}
0& V^1(\mmt,\l)& V^2(\mmt,\l)& V^3(\mmt,\l)\\
-V^1(\mmt,\l)&0& A^3(\mmt,\l)& -A^2(\mmt,\l)\\
-V^2(\mmt,\l)&-A^3(\mmt,\l)&0& A^1(\mmt,\l)\\
-V^3(\mmt,\l)&A^2(\mmt,\l)&-A^1(\mmt,\l)&0
\end{array}
\right), \quad \l =+1,0,-1.
\label{F-Ten}
\end{equation}
The explicit form of the polar and axial three-vectors are calculated as
\begin{align}\label{V-A-tensor}
\veg{ V}(\mmt,+1)&=\frac{{\mathcal N}}{\sqrt{2}}\left(
\begin{array}{c}
 p_0^2+m p_0-p_x p_{+} \\
 i \left(p_0^2+m p_0+i p_{y} p_{+}\right) \\
 -p_z p_{+}
\end{array}
\right),
&{\veg A}(\mmt,+1)=&\frac{{\mathcal N}}{\sqrt{2}}
\left(
\begin{array}{c}
 -i (m+p_0) p_z \\
 (m+p_0) p_z \\
 i(m+p_0) p_{+}
\end{array}
\right),\nonumber \\
\veg{ V} (\mmt,0 )&={\mathcal N}\left(
\begin{array}{c}
  p_x p_z \\
  p_y p_z \\
 -\left(p_0^2+m p_0-p_z^2\right)
\end{array}
\right),
&{\veg A}(\mmt,{0})=&{\mathcal N}
\left(
\begin{array}{c}
 - (m+p_0) p_y \\
  (m+p_0) p_x \\
 0
\end{array}
\right),\\
%%%%%%
\veg{ V}(\mmt,-1)&=\frac{{\mathcal N}}{\sqrt{2}}\left(
\begin{array}{c}
 -p_0^2-m p_0+p_x p_{-} \\
 i \left(p_0^2+m p_0-i p_y p_{+} \right) \\
  p_z p_{-}
\end{array}
\right),
&{\veg A}(\mmt,-1)=&\frac{{\mathcal N}}{\sqrt{2}}
\left(
\begin{array}{c}
 -i (m+p_0) p_z \\
 -(m+p_0) p_z \\
 i (m+p_0) p_{-}
\end{array}
\right),\nonumber
\end{align}
where $p_{\pm}=p_{x}\pm ip_{y}$, ${\mathcal N}=[m(m+p_0)]^{-1}$. The above states satisfy
\aeq
\frac{\veg{p}}{p_0}\times \veg{V}(\mmt,\l)=\veg{A}(\mmt,\l),\quad \veg{p}\cdot\veg{A}(\veg{p},\l)=0.
\ceq
Using these states we calculate the multipole moments for the current in Eq.(\ref{currentbv}). The gyromagnetic ratio,
here denoted by $\gva$, remains as usual unspecified, so far. The  electric and magnetic densities corresponding to the
current in (\ref{currentbv})  are labeled by (${\textit VA}$). The transverse densities are found  as,
\begin{subequations}\label{VAdensities1}
\aec
\varrho_E^{\textit VA}(\veg{q},1,\pm 1)&=&\frac{\charge((\veg{q}^2+\o^2)(q_x^2+q_y^2-2\veg{q}^2)-4m^2(q_x^2+q_y^2))}{8m^2(4m^2-\o^2)}\nonumber\\
&&-\frac{\charge \gva \veg{q}^2(q_x^2+q_y^2-2\veg{q}^2)}{4m^2(4m^2-\o^2)}\\
&=&\frac{\charge(4m^2-(\gva-1)(q_x^2+q_y^2+2q_z^2))}{4m^2},\\
\varrho_M^{\textit VA}(\veg{q},1,\pm 1)&=&\pm \frac{i \gva \charge q_z}{m}.
\cec
\end{subequations}
For the longitudinal polarization, $\l=0$, we encounter
\begin{subequations}\label{densities_LT}
\aec
\varrho_E^{\textit VA}(\veg{q},1,0)&=&\frac{\charge(4m^2(q_x^2+q_y^2-\veg{q}^2)-(\veg{q}^2+\o^2)(q_x^2+q_y^2))}{4m^2(4m^2-\o^2)}+\frac{\charge \gva \veg{q}^2(q_x^2+q_y^2)}{2m^2(4m^2-\o^2)}\\
&=&\frac{\charge(2m^2-(\gva-1)(q_x^2+q_y^2))}{2m^2},\\
\varrho_M^{\textit VA}(\veg{q},1,0)&=&0.
\cec
\end{subequations}
The three  resulting multipole moments are then calculated as,
\begin{eqnarray}
\left( Q_E^0\right)^{\textit{VA}}=\charge, \quad \left( Q_M^1\right)^{\textit VA}=\frac{\charge \gva}{2m}\langle S_z\rangle, \quad
\left( Q_E^2\right)^{\textit{VA}}=-\frac{\charge(1-\gva)}{m^2}\langle 3S_z^2-{\veg S}^2\rangle.
\label{bvct_emm}
\end{eqnarray}
The spin-1 multipole moments in (\ref{bvct_emm}) are same  as those of spin-1 residing in the four- vector
and given in Eqs.~(\ref{NKR_hfhf}) for $g_{\textit VA}=2$, {\it except for a relative sign between the electric quadrupole moments}.
This sign of $\left( Q_E^2\right)^{\textit VA}$ is inverse relative to that  of
$\left( Q_E^2\right)^{V}$  and technically reflects the opposite signs of $q_z^2$ in the combinations
$({\veg q}^2\mp q_z^2)$ determining $\varrho^V_{E}(1,1)$ in (\ref{NKR_MM1}), and
$\varrho^{\textit VA}_{E}(1,1)$  in (\ref{VAdensities1}), respectively.
The different dependencies of the electric densities in the $(1,0)\oplus(0,1)$ and $(1/2,1/2)$
representations on $q_i^2$ is of pure kinematic origin, and can be thought of
as a difference in the orientations of the four-vector and bi-vector quadrupoles in the Breit frame.
It definitively appears in consequence of  the different forms taken by  the boost in the respective
four-vector--, and bi-vector representations.
Setting $\gv =\gva =2$, and for the maximal polarization, our formalism predicts a negative
quadrupole moment of $Q_{E}^{2}=-e/m^2$ for the four-vector (same as the $W$ in the standard model) and a positive one,
$Q_{E}^{2}=+e/m^2$,  for the bivector. It is interesting to notice that a positive electric quadrupole moment, $Q_E^2=0.33 e/m^2$,
has been extracted from studying the $\rho$ meson in the light-front quark-model \cite{Jaus:2002sv,Cardarelli:1994yq}. This result,
and our findings are indicative of the possibility that the $\rho$ meson could transform in the bi-vector representation. An
interesting dynamical aspect of the map between the  anti-symmetric
Lorentz tensor  in (\ref{F-Ten}) and the Lorentz bi-vector in (\ref{bvmap})
is revealed upon associating the ${\mathcal F}^{\m\n}$ tensor  with the  $\rho$ meson and its dual (opposite parity) with the $a_1$ meson.
The bi-vector space is interesting as a carrier space for the $\rho $ meson in so far as it easily allows for a tensor coupling,
and is at variance to the  electroweak gauge  bosons, whose interactions with the matter fields are restricted by the principle
of minimal couplings. The
interactions of the non-gauge  $\rho$ and $a_1$ mesons with baryons can be of any type compatible with Lorentz
covariance and the symmetries of the light sector of QCD. The $\rho$ and $a_1$ mesons are well known to dominate
the respective vector- and axial vector hadron currents, they have  tensor couplings,  and are essential, among others, in the design of
chiral Lagrangians.  Within this context, the tensor electromagnetic currents of these mesons acquire importance.
Such hadronic versions of ${\mathcal F}^{\mu\nu}$ and its dual,  hint on the importance of the
$\rho$ and $a_1$  meson clouds surrounding the nucleon ($N$)
for the description of multipole moments in the transition of the nucleon to spin-3/2 resonances $(N^\ast$) such as the
$\Delta (1232)$ and $D(1520)$ resonances,
where the  $NN^\ast\gamma $  vertex (to be considered in a forthcoming work)  is
designed by a contraction of the electromagnetic field-strength tensor with a third-rank Lorentz tensor
sandwiched between the nucleon and the $N^\ast $ resonance state \cite{Jones:1972ky}.

\subsection{Spin-3/2 multipole moments}

The present section is devoted to the multipole decompositions of the currents in (\ref{gordonNKR}) and (\ref{RS_gordon}).
Again, if one wishes to calculate the spin-3/2 multipole moment, knowledge on the four-vector spinors is required.
We use the spin 3/2 basis previously employed in \cite{DelgadoAcosta:2009ic},
\begin{subequations}\label{base32}
\begin{eqnarray}
u^\alpha(\veg{p},3/2)&=&\eta^\alpha(\veg{p},1)u(\veg{p},1/2),\\
u^\alpha(\veg{p},1/2)&=&\frac{1}{\sqrt{3}}\eta^\alpha(\veg{p},1)u(\veg{p},-1/2)+\sqrt{\frac{2}{3}}\eta^\alpha(\veg{p},0)u(\veg{p},1/2),\\
u^\alpha(\veg{p},-1/2)&=&\frac{1}{\sqrt{3}}\eta^\alpha(\veg{p},-1)u(\veg{p},1/2)+\sqrt{\frac{2}{3}}
\eta^\alpha(\veg{p},0)u(\veg{p},-1/2),\\
u^\alpha(\veg{p},-3/2)&=&\eta^\alpha(\veg{p},-1)u(\veg{p},-1/2),
\end{eqnarray}
\end{subequations}
where $\overline{u}^\alpha(\veg{p},\lambda)\cdot u_\alpha(\veg{p},\lambda)=-1$ with $u(\veg{p},\l)$ and $\eta^\alpha (\veg{p},\lambda)$ being defined in (\ref{base12}) and (\ref{base1}) respectively.
\subsubsection{Rarita-Schwinger multipoles}
The calculation of the spin-3/2 multipole moments within the Rarita-Schwinger framework in (\ref{RS_gordon}) amounts to
the following  densities,
\begin{eqnarray}
\varrho_E^{RS}(\veg{q},3/2,\pm3/2)&=&\frac{\charge(4m^2-q_z^2+\veg{q}^2)(8m^2-(\gs-2)\veg{q}^2)}{16m^3 \o},\\
\varrho_M^{RS}(\veg{q},3/2,\pm 3/2)&=&\mp \frac{i \charge \gs q_z \left(2 m^2+q_z^2-\omega ^2\right)}{2 m^2 \o}.
\end{eqnarray}
For $\l=\pm1/2$ one has,
\begin{eqnarray}
\varrho_E^{RS}(\veg{q},3/2,\pm 1/2)&=&\frac{\charge(12m^2+3q_z^2+\veg{q}^2)(8m^2-(\gs-2)\veg{q}^2)}{48m^3 \o},\\
\varrho_M^{RS}(\veg{q},3/2,\pm 1/2)&=&\pm \frac{i \charge \gs q_z \left(22 m^2+9q_z^2-5\omega ^2\right)}{6 m^2 \o}.
\end{eqnarray}
The general expressions for any polarization in the notations of (\ref{simpnot}) and for the maximal polarization
are then
\begin{subequations}\label{raripoles}
\aec
\left( Q_E^0\right)^{RS}&=&\charge,\\
\left( Q_M^1\right)^{RS}&=&\frac{\charge \gs}{2m}\frac{1}{3}\langle S_z\rangle,\\
\left( Q_E^2\right)^{RS}&=&\frac{\charge }{m^2}\frac{1}{3}\langle \,{\mathcal A}\,\rangle,\\
\left( Q_M^3\right)^{RS}&=&\frac{\charge \gs}{2m^3}\langle\,{\mathcal B}\,\rangle.
\cec
\end{subequations}
The operators ${\mathcal A}$ and ${\mathcal B}$  in the four-vector spinor representation read,
\begin{subequations}\label{ayb}
\begin{eqnarray}
{\mathcal A}=3\,S_z^2-\veg{S}^2,&\quad&
{\mathcal B}=S_z\(15S_z^2-\frac{41}{5}\veg{S}^2\).
\end{eqnarray}
\end{subequations}
One sees that it is exclusively the Dirac spin-tensor which determines the dipole and octupole magnetic moments,
without contributing to the electric quadrupole moment. The spin-tensor of the underlying spin-1 sector does not show up nowhere.

%-----------------------------------------------------------------------------------------------------------------
\subsubsection{Non-Lagrangian currents and  multipoles for the four-vector-spinor}
%---------------------------------------------------------------------------------------------------------------
In order to gain a deeper insight into the problematics  of eq.~(\ref{raripoles}), it is quite instructive to
compare  (\ref{RS_gordon}) to the  most general form of a covariant  current density, satisfying  time-reversal--,
parity-,  and gauge invariance for the $\rsrep$ representation, and
given by \cite{Nozawa:1990gt},
\aeq
j^{a_1,a_2,b_1,c_1,c_2}_\m(\mmt', \lambda^\prime;\mmt,\lambda )=-\,\charge\, \overline{u}^\a(\mmt', \lambda^\prime)\,
V_{\a\b\m}(p',p;a_1,a_2,b_1,c_1,c_2)\,u^\beta (\mmt, \lambda ).
\label{gencur}
\ceq
Here
\begin{align}
V_{\a\b\m}(p',p;a_1,a_2,b_1,c_1,c_2)\,=\,
&a_1 2m \g_\m g_{\a\b}+a_2(p'+p)_\m g_{\a\b}\nonumber\\
&+b_1\((p'-p)_\a g_{\m\b}-(p'-p)_\b g_{\m\a}\)\nonumber\\
&+\frac{1}{(2m)^2}(p'-p)_\a(p'-p)_\b\(2m c_1 \g_\m+c_2(p'+p)_\m\).
\end{align}
The multipole moments resulting from the latter non-Lagrangian  current density are calculated as,
\begin{subequations}\label{Nozawa}
\begin{align}
Q_E^0&= \charge(a_1+a_2),\\
Q_M^1&=\frac{\charge}{2m}\frac{2}{3}(a_1+b_1)\bra S_z\ket,\\
Q_E^2&=\frac{\charge}{m^2}\frac{1}{6}(2a_1+2a_2-2b_1-c_1-c_2)\bra \,{\mathcal A}\,\ket,\\
Q_M^3&=\frac{\charge}{2m^3}(2a_1-c_1)\bra \,{\mathcal B}\,\ket.
\end{align}
\end{subequations}
The Rarita-Schwinger Lagrangian which is linear in the momentum corresponds to setting all parameters equal to zero except  $a_1$. This inevitably
restricts the parameters in $j^{a_1,a_2,b_1,c_1,c_2}_\m(\mmt', \lambda^\prime;\mmt,\lambda )$
to
\aeq\label{rsp}
a_1=\frac{g_S}{2},\qquad a_2\,=\,-\frac{g_S}{2}+1,\,\qquad b_1\,=\,c_1\,=\,c_2\,=\,0, \quad g_S=2.
\ceq
Instead, the NKR current,
is recovered for the following parameter set,
\aeq\label{cpfp}
a_1=\frac{g}{2},\qquad a_2\,=\,-\frac{g}{2}+1,\,\qquad b_1\,=\,g,\qquad c_1\,=\,c_2\,=\,0,
\ceq
to be inserted in Eqs.~(\ref{Nozawa}). In effect, the essential difference between the NKR and the RS spin-3/2 multipole moments
in the respective Eqs.~(\ref{cpfp}), and  (\ref{rsp}), is brought about by the non-vanishing value of the $b_1$ parameter.
The $b_1$ dependent term in (\ref{gencur}) takes its origin  from the vector part of the generators of the $\rsrep$ representation
and  is completely missing from the Rarita-Schwinger formalism. A $b_1=0$ value  is responsible for the absence of a
contribution to the magnetic dipole moment by the vector part of $\psi_\mu$, and for  the $g\not=2 $
value in the Rarita-Schwinger framework.
This shortcoming has been corrected  in \cite{Lorce:2009bs},
where a non-Lagrangian current has been placed on the light cone (LC).
The LC moments reported in  \cite{Lorce:2009bs} would correspond to the following
$j^{a_1,a_2,b_1,c_1,c_2}_\m(\mmt', \lambda^\prime;\mmt,\lambda )$ parametrization,
\begin{equation}
a_1=3, \quad a_2=-2, \quad b_1=0, \quad  c_1=8, \quad c_2=0,
\end{equation}
and are listed in Table I. We see that $c_1\neq 0$, in consequence of which the current becomes second order in the momenta.

Our point here is that the multipole moments of this quadratic in the momenta current can be equally well obtained also
from a current linear in the momentum. It is straightforward to verify that such a current would correspond to the following
parametrization:
\begin{equation}
a_1=-1, \quad a_2=2, \quad b_1=4, \quad  c_1=c_2=0,
\end{equation}
It translates into the following three-term linear current,
\begin{eqnarray}
j_\m(\mmt', \lambda^\prime;\mmt,\lambda )\vert_{LC}
&=&-\charge\,\overline{u}^{\alpha}(\mmt',\lambda^\prime)
\[ g_{\alpha\beta}(p'+p)_\mu+i \(-2M^S_{\m\n}g_{\a\b}+4[M^{V}_{\mu\nu}]_{\alpha\beta}\)(p'-p)^\nu\]u^\beta(\mmt,\lambda ),
\nonumber\\
\label{Cedric_32}
\end{eqnarray}
which mimics the second-order Light Cone current constructed by Lorc\'e in \cite{Lorce:2009bs}. Subsequently,
$j_\m(\mmt', \lambda^\prime;\mmt,\lambda )\vert_{LC}$ will be termed to as Lorc\'e-like linear current.
The above considerations show that the non-Lagrangian spin-3/2  currents of second order
are not really inevitable  for the explanation of the multipole moments. The  current in (\ref{Cedric_32}) excludes equally well the
coupling of the spin-1/2 sector from $\psi_\mu$ to the electromagnetic field, though it ceases to be  a genuine
Poincar\'e covariant projector current due to the inadequate combination between the spin-magnetization tensors of the spinor- and four-vector
sectors of $\psi_\mu$. We here tested the Lorc\'e-like current in (\ref{Cedric_32}) in the calculation of Compton scattering
off a spin-3/2 target. The essentials of the calculation are given in the Appendix.
We observe that the tree-level forward Compton scattering cross corresponding to the current in (\ref{Cedric_32})
grows infinitely with the energy according to,
\aeq
\[\frac{d \s}{d \Omega}\]_{\q=0}=\frac{1}{81}(4712\h^4-16\h^2+81)r_0^2,
\label{Cedric_Cmpt32}
\ceq
Here $r_0=e^2/(4\pi m)=\a/m$,  and $\eta =\omega /m$, where $\omega$ and $m$ stand in turn for the energy of the incident photon,
and the mass of the target.
This contrasts the high-energy forward scattering cross section concluded from the genuinely NKR  spin-3/2  current
in (\ref{gordonNKR}), in which case one finds finite values according to \cite{DelgadoAcosta:2009ic},
\begin{equation}
\[\frac{d \s}{d \Omega}\]_{\q=0}=\frac{1}{24}\left((g-2)^4\eta^2 +24\right)r_0^2.
\label{German_Cmpt32}
\end{equation}
We conclude that non-linear non-Lagrangian spin-3/2 currents are not necessarily more general and more advantageous 
than the linear spin-3/2  Lagrangian current emerging within the covariant projector formalism. Furthermore, a Lagrangian 
method is always more advantageous for field-theoretical considerations. The above observation  does not exclude by no 
means the possibility that Compton scattering within Lorc{\'e }'s Light-Cone  framework  \cite{Lorce:2009bs}, and under 
employment of the genuine second order Lorc{\'e} current, may throughout create a result  consistent with unitarity.

%-------------------------------------------------------------------------------------------------------------
\subsubsection{Covariant-projector multipoles }
%--------------------------------------------------------------------------------------------------------------
The transverse spin 3/2 NKR  densities of the current in (\ref{gordonNKR}) for the maximal polarizations
 are calculated as:
\begin{eqnarray}
\varrho_E(\veg{q},3/2,\pm 3/2)&=&\frac{\charge \omega  \left(\omega ^2-q_z^2\right)}{8 m^3}+\frac{\gsv \charge \left(-3 \omega ^4+3 \left(4 m^2+q_z^2\right) \omega ^2-4 m^2 q_z^2\right)}{16 m^3 \omega },\\
\varrho_M(\veg{q},3/2,\pm 3/2)&=&\mp \frac{i \gsv q_z \charge \left(2 m^2+q_z^2-2 \omega ^2\right)}{2 m^2 \omega }.
\end{eqnarray}
For $\l=\pm 1/2$ one finds,
\aec
\varrho_E(\veg{q},3/2,\pm 1/2)&=&\frac{\charge \omega  \left(8 m^2+3 q_z^2
+\omega ^2\right)}{24 m^3}\nonumber\\
&+&\frac{\gsv \charge \left(32 m^4+4 \left(3 q_z^2+\omega ^2\right) m^2-3 \left(\omega ^4+3 q_z^2 \omega ^2\right)\right)}{48 m^3 \omega
   },\\
\varrho_M(\veg{q},3/2,\pm 1/2)&=&\pm \frac{i \gsv q_z \charge \left(22 m^2+9 q_z^2-4 \omega ^2\right)}{6 m^2 \omega }.
\end{eqnarray}
The electromagnetic moments for the above densities are then found as
\begin{subequations}
\begin{eqnarray}
(Q_E^0)^{NKR}&=&\charge,\\
(Q_M^1)^{NKR}&=&\frac{\charge \gsv}{2m}\langle S_z\rangle,\\
(Q_E^2)^{NKR}&=&\frac{\charge(1-\gsv)}{m^2}\frac{1}{3}\langle \,{\mathcal A}\,\rangle,\\
(Q_M^3)^{NKR}&=&\frac{\charge \gsv}{2m^3}\langle\,{\mathcal B}\,\rangle,
\end{eqnarray}
\label{fvsmultipoles}
\end{subequations}
again with the use of (\ref{simpnot}) for $s=3/2$ and $\l=\pm 1/2,\pm 3/2$.

%_________________________________________________________________________________________________________________________________
\subsection{Spin $3/2$ multipole moments in the $\tmrep$ representation}
%__________________________________________________________________________________________________________________________________
The current for representations of this type is given in Eq.(\ref{jcurrent}) and for the  $j=3/2$ value of interest reads,
\aeq
j^{(3/2)}_\mu(\mmt',\l';\mmt,\l)=\charge\,\overline{w}(\mmt',\l')\left[ (p'+p)_\m + ig  M^{(3/2)}_{\m\n}(p'-p)^\n\right]w(\mmt,\lambda),
\label{tmcurrent}
\ceq
where we have simply denoted by $g$ the corresponding gyromagnetic factor. The rotational generators in this single spin-3/2 representation satisfy
$(\mathbf{J}\cdot\mathbf{n})^{4}=\frac{5}{2}(\mathbf{J}\cdot\mathbf{n})^{2}-(\frac{3}{4})^2$,
a relationship that we employ in the calculation of the boost elements,
\begin{align}
\exp (\pm\mathbf{J}\cdot\mathbf{n})\varphi &=\cosh\frac{\varphi}{2} \left(1-\frac{1}{2}\sinh^2\frac{\varphi}{2}\right)
\pm (\mathbf{J}\cdot\mathbf{n})\sinh\frac{\varphi}{2}\left(2-\frac{1}{3}\sinh^2\frac{\varphi}{2}\right) \nonumber \\
&+ 2(\mathbf{J}\cdot\mathbf{n})^{2}\cosh \frac{\varphi}{2} \sinh^2\frac{\varphi}{2}
\pm \frac{4}{3}(\mathbf{J}\cdot\mathbf{n})^{3}\sinh^3\frac{\varphi}{2}.
\end{align}
This relation can be used to explicitly calculate the boost generator in Eq. (\ref{boost32}). The states of well defined parity
in any frame are constructed by boosting the corresponding rest frame states.
We aim to compare  the spin-3/2$^-$ multipole moments of  the four-vector-spinor,
and focus on the four negative parity states,
\begin{subequations}\label{base320032}
\begin{eqnarray}
v(\veg{p},+3/2)&=&\frac{1}{4(m(m+p_0))^{3/2}}\left(
\begin{array}{c}
 (m+p_0+p_z)^3 \\
 \sqrt{3} p_+ (m+p_0+p_z)^2 \\
 \sqrt{3} p_+^2 (m+p_0+p_z) \\
 p_+^3 \\
 -(m+p_0-p_z)^3 \\
 \sqrt{3} p_+ (m+p_0-p_z)^2 \\
 -\sqrt{3} p_+^2 (m+p_0-p_z) \\
 p_+^3
\end{array}
\right),\\
v(\veg{p},+1/2)&=&\frac{1}{4(m(m+p_0))^{3/2}}\left(
\begin{array}{c}
 \sqrt{3} p_- (m+p_0+p_z)^2 \\
 -(m+p_0+p_z) \left(3 p_z^2+(m-3 p_0) (m+p_0)\right) \\
 p_+ \left((m+p_0) (m+3 p_0)-3 p_z^2\right) \\
 \sqrt{3} p_+^2 (m+p_0-p_z) \\
 \sqrt{3} p_- (m+p_0-p_z)^2 \\
 (m+p_0-p_z) \left(3 p_z^2+(m-3 p_0) (m+p_0)\right) \\
 p_+ \left((m+p_0) (m+3 p_0)-3 p_z^2\right) \\
 -\sqrt{3} p_+^2 (m+p_0+p_z)
\end{array}
\right),\\
v(\veg{p},-1/2)&=&\frac{1}{4(m(m+p_0))^{3/2}}\left(
\begin{array}{c}
 \sqrt{3} p_-^2 (m+p_0+p_z) \\
 p_- \left((m+p_0) (m+3 p_0)-3 p_z^2\right) \\
 -(m+p_0-p_z) \left(3 p_z^2+(m-3 p_0) (m+p_0)\right) \\
 \sqrt{3} p_+ (m+p_0-p_z)^2 \\
 -\sqrt{3} p_-^2 (m+p_0-p_z) \\
 p_- \left((m+p_0) (m+3 p_0)-3 p_z^2\right) \\
 (m+p_0+p_z) \left(3 p_z^2+(m-3 p_0) (m+p_0)\right) \\
 \sqrt{3} p_+ (m+p_0+p_z)^2
\end{array}
\right),\\
v(\veg{p},-3/2)&=&\frac{1}{4(m(m+p_0))^{3/2}}\left(
\begin{array}{c}
 p_-^3 \\
 \sqrt{3} p_-^2 (m+p_0-p_z) \\
 \sqrt{3} p_- (m+p_0-p_z)^2 \\
 (m+p_0-p_z)^3 \\
 p_-^3 \\
 -\sqrt{3} p_-^2 (m+p_0+p_z) \\
 \sqrt{3} p_- (m+p_0+p_z)^2 \\
 -(m+p_0+p_z)^3
\end{array}
\right),
\end{eqnarray}
\end{subequations}
where $p_{\pm}=p_x\pm ip_y$. The transverse electric densities for the maximal polarizations obtained with these states are:

\begin{eqnarray}
\varrho_E\(\frac{3}{2},\pm\frac{3}{2}\)&=&-\frac{\charge \omega  \left(12 m^2+3 q_x^2+3 q_y^2-4 \omega ^2\right)}{8 m^3}\nonumber\\
&+&\frac{3 \charge g\left(-16 m^4-4 \left(q_x^2+q_y^2-5 \omega ^2\right) m^2-4 \omega ^4+3 \left(q_x^2+q_y^2\right) \omega
   ^2\right)}{16 m^3 \omega },\\
\varrho_E\(\frac{3}{2},\pm\frac{1}{2}\)&=&\frac{\charge \left(4 m^2+3 \left(q_x^2+q_y^2\right)\right) \omega }{8 m^3}\nonumber\\
&+&\frac{\charge g\left(16 m^4+4 \left(3 q_x^2+3 q_y^2-\omega ^2\right) m^2-9 \left(q_x^2+q_y^2\right) \omega ^2\right)}{16 m^3
   \omega },
\end{eqnarray}
where $\o^2=\veg{q}^2-4m^2$. The magnetic densities are calculated as
\begin{align}
\varrho_M\(\pm\frac{3}{2},\pm\frac{3}{2}\)&=\pm\frac{3 i \charge g q_z \left(2 m^2+q_z^2\right)}{2 m^2 \omega }\\
\varrho_M\(\pm\frac{3}{2},\pm\frac{1}{2}\)&=\mp\frac{i \charge g q_z \left(-2 m^2-6 q_x^2-6 q_y^2+3 q_z^2\right)}{2 m^2 \omega }.
\end{align}
From these densities we find the following multipole moments for a particle transforming in the $\tmrep$ representation
\begin{subequations}\label{set320032}
\begin{eqnarray}
Q_E^0&=&\charge,\\
Q_M^1&=&\frac{\charge g}{2m}\langle S_z\rangle,\\
Q_E^2&=&-\frac{\charge(1-g)}{m^2}\langle \,{\mathcal A}\,\rangle,\\
Q_M^3&=&-\frac{\charge g}{2m^3}3\langle\,{\mathcal B}\,\rangle,
\end{eqnarray}
\end{subequations}
with the operators ${\mathcal A}$ and ${\mathcal B}$ defined in Eq. (\ref{ayb}). We observe that according to  the covariant
projector formalism,  all multipoles of a state transforming in the $\tmrep$ are dictated by the value
of a single parameter, the gyromagnetic factor $g$. Furthermore, comparing with the expressions in Eq.(\ref{fvsmultipoles})
for the four-vector-spinor  we see that only the charge and magnetic moment coincide. Higher multipole
moments turn out to be representation specific.
This is so because magnetic dipole moments, in describing the rest-frame coupling to the magnetic field
are exclusively determined by the generators of rotations in the representation of interest,
which are necessarily same for equal spins. This is no longer valid for higher multipoles,
which are sensitive to the dependence of the boost operator on the momentum, a dependence
which  varies with the representation.

\section{Summary and discussion}

In the Table \ref{table1} below all the results obtained so far have been summarized.

\begin{table}[ht]
\centering          % used for centering table
\begin{tabular}
[c]{|c|c|c|c|c|c|c|c|}\hline\hline
\multicolumn{1}{||c|}{Formalism} & \multicolumn{1}{||c|}{Representation} &
\multicolumn{1}{||c|}{Spin} & \multicolumn{1}{||c|}{$g$-factor} &
\multicolumn{1}{||c|}{$Q_{E}^{0}$} & \multicolumn{1}{||c|}{$Q_{M}^{1}$} &
\multicolumn{1}{||c|}{$Q_{E}^{2}$} & \multicolumn{1}{||c||}{$Q_{M}^{3}$%
}\\\hline\hline
Dirac & $(\frac{1}{2},0)\oplus(0,\frac{1}{2})$ & $\frac{1}{2}$ & $g_{D}=2$ &
$e$ & $\frac{eg_{D}}{2m}\langle S_{z}\rangle$ & $0$ & $0$\\\hline
NKR & $(\frac{1}{2},0)\oplus(0,\frac{1}{2})$ & $\frac{1}{2}$ & $g_{S}=2$ & $e$
& $\frac{eg_{S}}{2m}\langle S_{z}\rangle$ & $0$ & $0$\\\hline
Proca & $(\frac{1}{2},\frac{1}{2})$ & $1$ & $g_{P}=1$ & $e$ & $\frac{eg_{P}%
}{2m}\langle S_{z}\rangle$ & $0$ & $0$\\\hline
SM & $(\frac{1}{2},\frac{1}{2})$ & $1$ & $g_{W}=2$ & $e$ & $\frac{eg_{W}}%
{2m}\langle S_{z}\rangle$ & $-\frac{e(g_{W}-1)}{m^{2}}\langle\mathcal{A}%
\rangle$ & $0$\\\hline
LC & $(\frac{1}{2},\frac{1}{2})$ & $1$ & $g_{LC}=2$ & $e$ & $\frac{eg_{LC}}%
{2m}\langle S_{z}\rangle$ & $-\frac{e(g_{LC}-1)}{m^{2}}\langle\mathcal{A}%
\rangle$ & $0$\\\hline
NKR & $(\frac{1}{2},\frac{1}{2})$ & $1$ & $g_{V}=2$ & $e$ & $\frac{eg_{V}}%
{2m}\langle S_{z}\rangle$ & $-\frac{e(g_{V}-1)}{m^{2}}\langle\mathcal{A}%
\rangle$ & $0$\\\hline
NKR & $(1,0)\oplus(0,1)$ & $1$ & $\gva$ & $e$ & $\frac{e\gva}{2m}\langle
S_{z}\rangle$ & $\frac{e(\gva-1)}{m^{2}}\langle\mathcal{A}\rangle$ &
$0$\\\hline
RS & $(\frac{1}{2},\frac{1}{2})\otimes(\frac{1}{2},0)\oplus(0,\frac{1}{2})$ &
$\frac{3}{2}$ & $\frac{g_{D}}{3}=\frac{2}{3}$ & $e$ & $\frac{g_{D}}{3}\frac
{e}{2m}\langle S_{z}\rangle$ & $\frac{e}{m^{2}}\frac{1}{3}\langle
\mathcal{A}\rangle$ & $\frac{eg_{D}}{2m^{3}}\frac{1}{3}\langle\mathcal{B}%
\rangle$\\\hline
LC  & $(\frac{1}{2},\frac{1}{2})\otimes(\frac{1}{2}%
,0)\oplus(0,\frac{1}{2})$ & $\frac{3}{2}$ & $g_{LC}=2$ & $e$ & $\frac{e g_{LC}}%
{2m}\langle S_z \rangle $ & -$\frac{e}{m^{2}}\langle {\mathcal A}\rangle$
& $-\frac{e}{2m^{3}}\langle  2{\mathcal B}\rangle $\\\hline
NKR & $(\frac{1}{2},\frac{1}{2})\otimes(\frac{1}{2},0)\oplus(0,\frac{1}{2})$ &
$\frac{3}{2}$ & $\gsv=2$ & $e$ & $\frac{e\gsv}{2m}\langle S_{z}\rangle$
& $-\frac{e(\gsv-1)}{m^{2}}\frac{1}{3}\langle\mathcal{A}\rangle$ & $\frac{e\gsv%
}{2m^{3}}\langle\mathcal{B}\rangle$\\\hline
NKR & $(\frac{3}{2},0)\oplus(0,\frac{3}{2})$ & $\frac{3}{2}$ & $g$ & $e$ &
$\frac{eg}{2m}\langle S_{z}\rangle$ & $\frac{e(g-1)}{m^{2}}\langle
\mathcal{A}\rangle$ & $-\frac{eg}{2m^{3}}3\langle\mathcal{B}\rangle$\\\hline
\end{tabular}
\caption{Summary of the multipole moments. The abbreviations SM, RS, LC, and NKR stand in their turn
for Standard Model, Rarita-Schwinger, Light Cone \cite{Lorce:2009bs},
and the second order formalism of \cite {Napsuciale:2006wr}. For other notation,
see the main body of the text.
Notice that both the LC and NKR predictions on the spin 1 multipoles are in accord with the Standard Model.
}% title of Table
\label{table1}%
\end{table}

To recapitulate, we wish to emphasize the following two  main points reflected by the above results.
The first one concerns the incompleteness of the Proca and Rarita-Schwinger formalisms,
which becomes detectable in the former case through  a vanishing electric quadrupole moment, a result
which is a consequence of the inbuilt deficient  $g_{P}=1$ value, instead of the universal $g=2$ value.
In the latter case, the deficient $g_{RS}=\frac{2}{3}$ reflects the omission of the coupling to
the magnetic field of the  spin-magnetization current
of the vector sector of the four-vector spinor, an issue explained in
subsection {\bf D} of section II  above.
This very omission is further more at the root of the violation of unitarity
in Compton scattering in the forward direction \cite{DelgadoAcosta:2009ic}.
The second point concerns  the representation dependence of the multipole moments higher than the
charge monopole and the magnetic dipole of particles of equal spins transforming in different Lorentz
representations.

This result might seem surprising but it is quite natural indeed once we recall that
\begin{itemize}
\item the magnetic dipole moment, in describing the rest frame coupling to the magnetic field,
exclusively invokes the generators of rotations. In the multipole expansion, this moment is dictated
by the linear terms in the momentum expansion of the magnetic current density.
The linear terms are already contained in the operators of the electromagnetic currents, hence this moment
is insensitive to the momentum dependence of the states which compose the electromagnetic current.
\item the momentum dependence of the boost operator varies with the representation. This is reflected in a
different momentum dependence of the corresponding magnetic--, and electric-current densities. Higher multipoles
are dictated by the quadratic and higher terms in the momentum expansion of these current densities and are
sensitive to the momentum dependence of the states which is different for different representations.
\end{itemize}

\section{Conclusions}

In the present investigation we studied the electromagnetic multipole moments of spin-1/2, spin-1, and spin-3/2 particles
within the Poincar\'e covariant second order formalism of  \cite {Napsuciale:2006wr} and compared our results with those of
the Dirac, Proca, Standard Model and Rarita-Schwinger theories. Introducing the electromagnetic interactions via the gauge
principle, we were able to show that in the scheme under discussion we reproduce for spin-1/2 the couplings and
multipole moments known from the Dirac theory. Concerning spin-1, we studied two distinct Lorentz representations, the
four-vector and the bi-vector. As for the former, we explicitly showed  that Proca's theory is incomplete in the sense that it
yields vanishing electric quadrupole as a result of the insufficient prediction for the gyromagnetic factor as $g_{P}=1$. Then
predictions of the Poincar\'e covariant
formalism on the electromagnetic multipole moments of the four-vector representation coincided with the tree-level predictions of the
standard model. This achievement is not exclusive to the Poincar{\'e} covariant projector formalism.
More recently,  the standard model properties of the $W$ boson have been also successfully reproduced  by Lorc{\'e} \cite{Lorce:2009bs}
within a light-cone framework.
We furthermore  found that the electric-charge-- and magnetic-dipole moments of the bi-vector representation coincide with
those of the four-vector, while the electric quadrupole moment came out  of equal absolute value  but of opposite sign.

Regarding spin-3/2, we showed that the linear  Rarita-Schwinger framework can be obtained from the second order
Poincar\'e projector only on-shell. This linearization is the main culprit for the decoupling of
the four-vector building block  from  the
electromagnetic current, which happens to  capture  only  contributions from the underlying spinor sector.
The principal advantage of the second order  Poincar\'e covariant projector  formalism over
the linear  Rarita-Schwinger framework consists in the incorporation on equal footings of the vectorial and spinorial
building blocks of $\psi_\mu$ into the expression for the current.  In this fashion, the spin-3/2 multipole moments
are no longer determined by an incomplete $g=\frac{2}{3}$  but appear interrelated by a full-flashed spin-3/2  gyromagnetic factor
that takes the universal value of $\gsv=2$ and is in accord with unitarity of forward Compton scattering in the ultraviolet.

We compared in same representation our results with those following  from a particular non-Lagrangian current,
allowed to be second order in the momenta,  and constructed its counterpart within the covariant projector formalism, i.e. we found
its linear current equivalent of identical multipole  moments.
We concluded that, at least as it concerns the electromagnetic multipoles,
the non-Lagrangian spin-3/2 currents of higher order
in the momenta are not necessarily  more general and more advantageous than the
two-term spin-3/2  current emerging within the covariant projector method.

We also worked out the predictions of the Poincar\'e projector formalism for the multipoles of a particle transforming in
the single-spin $\tmrep$ representation  finding similar results as in the spin-1 case, i.e., charge and magnetic moments
coincide with those of the spinor-vector representation but the higher multipole moments (electric quadrupole and magnetic octupole)
differ, this time both in sign and in magnitude.

We furthermore observed  that if the electric quadrupole  moments of the
$\rho$ meson and the $W$ boson were to be of opposite signs, as suggested by light-front quark-model calculations
\cite{Jaus:2002sv,Cardarelli:1994yq}, then the relativistic  wave function of a fundamental  $\rho$ meson polarized in
$\z$ direction  has necessarily to transform as the vector field of a totally anti-symmetric tensor,
${\mathcal F}^{\mu\nu}_{(\z)}$. The axial-vector companion to the $\rho$ meson in ${\mathcal F}^{\mu\nu}_{(\z)}$ is described by the
dual tensor and carries the quantum numbers of the $a_1$ meson. This observation is compatible with the $\rho$-- and $a_1$ vector
meson dominance hypothesis of the respective strong vector--, and axial-vector currents.

Finally, we like to emphasize that the spin-3/2  description within the Poincar{\'e} covariant projector method is based on minimal gauging of a
second order Lagrangian, thus avoiding the commonly used {\it ad hoc} extension of the linear Rarita-Schwinger Lagrangian
by terms describing non-minimal couplings \cite{Ferrara}. The distinction between  minimal and  non-minimal couplings is best visualized within the
geometric interpretation of gauge theories, where the former stand for parallel transport, while the latter would invoke rare torsion couplings to the electromagnetic field.

\section*{Acknowledgments}
Work partly supported by CONACyT under project 156618.

%-------------------------------------------------------------------------------------------------
\section{Appendix: Forward Compton Scattering for non-Lagrangian light-cone electromagnetic moments}
Here we highlight the calculation of Compton scattering off a spin-3/2 target  using the linear Lorc\'e-like
current in (\ref{Cedric_32}), constructed within the NKR framework as a current having same multipole moments as
the genuine second order Lorc{\'e} current in \cite{Lorce:2009bs}. The linear Lorc{\'e}-like--, and NKR currents are
characterized by equal charge- and magnetic dipole moments and are distinct through   the higher multipoles.
The aim is to figure out whether the above difference will show up in the evaluation of the process under investigation.

The calculation is executed along the line of ref.~ \cite{DelgadoAcosta:2009ic}, however, we are using a
differently parametrized $\G_{\a\b\m\n}$ tensor in (\ref{gammageneral}).
Namely, we fixed those parameters in such a way  that the electromagnetic multipole moments
associated with the Lorc\'e-like current (\ref{Cedric_32}) are reproduced. This is achieved by the following replacements,
\aeq
\frac{1}{2}(2f+g_S)\rightarrow -1,\quad -\frac{1}{2}(2f+g_S-2)\rightarrow +2,\quad -\frac{1}{3}(3 f+3g_V+1)\rightarrow -4,
\ceq
in combination  with (\ref{cfdf}), with the purpose of  eliminating  spin- 3/2 $\leftrightarrow$ spin- 1/2 transitions.

From the corresponding  second order interaction Lagrangian, one can extract the Feynman rules needed for constructing
the tree-level Compton scattering amplitudes, and use as a propagator the inverse of the equation of motion:
\begin{align}
\P_{\a\b}(p)=&(\G_{\a\b\m\n}p^\m p^\n-m^2g_{\a\b})^{-1}\nonumber\\
=&\frac{1}{p^2-m^2+i\e}\[ P^{(3/2)}_{\a\b}(p)-\frac{p^2-m^2}{m^2}P_{\a\b}^{(1/2)}(p)\].
\end{align}
Here $P^{(1/2)}_{\a\b}(p)$ and $P^{(3/2)}_{\a\b}(p)$ are the respective  spin-1/2 and spin-3/2 projectors,
\begin{align}
P^{(1/2)}_{\a\b}(p)=&\frac{1}{3}\g_\a\g_\b+\frac{1}{3p^2}(\not{p}\g_\a p_\b+p_\a\g_\b\not{p}),\\
P^{(3/2)}_{\a\b}(p)=&g_{\a\b}-\frac{1}{3}\g_\a\g_\b-\frac{1}{3p^2}(\not{p}\g_\a p_\b+p_\a\g_\b\not{p}).
\end{align}
The averaged squared amplitude is then found as:
\aeq
\overline{\vert\M\vert^2}=\frac{1}{8}\sum_{pol}\vert\M\vert^2=\frac{e^4}{8}\Tr[P^{\h\a}(p')U_{\a\b\m\n}P^{\b\z}(p)\overline{U}^{\z\h\m\n}],
\ceq
with:
\begin{align}
U_{\m\n}=&V(p',Q)_{\a\g\m} \P^{\g\d}(Q)V(Q,p)_{\d\b\n}+V(p',R)_{\a\g\n}\P^{\g\d}(R)V(R,p)_{\d\b\m}+V_{\a\b\m\n},\\
\overline{U}_{\m\n}=&V(p,Q)_{\z\f\n} \P^{\f\q}(Q)V(Q,p')_{\q\h\m}+V(p,R)_{\z\f\m}\P^{\f\q}(R)V(R,p')_{\q\h\n}+V_{\z\h\m\n},
\end{align}
where $Q=p+q=p'+q'$, and $R=p'-q=p-q'$ are in turn the momenta of the intermediate states in the $s$ and $u$ channels,
$p\, (p')$ stands for the momentum of the incident (scattered) fermion, while $q\,(q')$ is the momentum of the incident (scattered) photon.
 The first and second order vertices are given by
\begin{align}
V(p',p)_{\a\b\m}&=\G_{\a\b\n\m}p'^\n+\G_{\a\b\m\n}p^\n,\\
V_{\a\b\m\n}&=-(\G_{\a\b\m\n}+\G_{\a\b\n\m}),
\end{align}
and satisfy the Ward-Takahashi identity:
\aeq {\label{ward}}
(k'-k)^\m V(k',k)_{\a\b\m}=\P_{\a\b}^{-1}(k')-\P_{\a\b}^{-1}(k).
\ceq
The latter relation ensures gauge invariance of the scattering amplitude and as we can see, the r.h.s of (\ref{ward}) is independent of
the undetermined parameters, meaning the Ward-Takahashi identity holds for any parametrization in the second order formalism.

%%%%%%%%%%%%%%%%%%%%%%%%%%%%%%%%%%%%%%%%%%%%%%%%%%%%%%%%%%%%%%%%%%%%%%%%%%%%%%%%%%%%%%%%%%%%%%%%%%%
%%%%%%%%%%%%%%%%%%%%%%%%%%%%%%%%%%%%%%%%%%%%%%%%%%%%%%%%%%%%%%%%%%%%%%%%%%%%%%%%%%%%%%%%%%%%%%%%%%%
\begin{figure}[ht]\begin{center}
\includegraphics[scale=0.75]{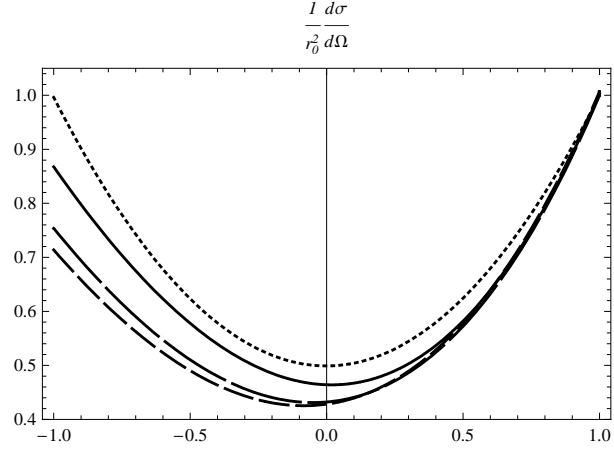}
\end{center}
\caption{\label{low} Differential cross section for Compton scattering off spin 3/2 particles at $\h=0.1$. The dotted line corresponds to the
Thompson classical limit, the short-dashed line represents the Rarita-Schwinger result, the long-dashed line indicates the prediction
based on the Lorc\'e-like first order current.
The continuous line refers to the prediction by the NKR method.}
\end{figure}
%%%%%%%%%%%%%%%%%%%%%%%%%%%%%%%%%%%%%%%%%%%%%%%%%%%%%%%%%%%%%%%%%%%%%%%%%%%%%%%%%%%%%%%%%%%%%%%%%%%
%%%%%%%%%%%%%%%%%%%%%%%%%%%%%%%%%%%%%%%%%%%%%%%%%%%%%%%%%%%%%%%%%%%%%%%%%%%%%%%%%%%%%%%%%%%%%%%%%%%
\begin{figure}[ht]\begin{center}
\includegraphics[scale=0.75]{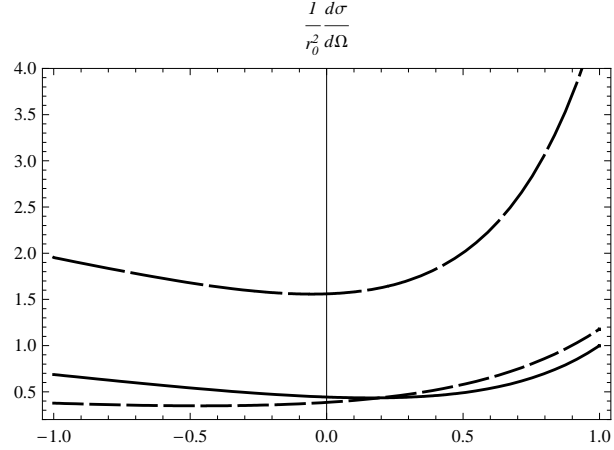}
\end{center}
\caption{\label{high} Differential cross section for Compton scattering off spin 3/2 particles as a function of $x=\cos\q$ at $\h=0.5$,
for the Rarita-Schwinger theory (short-dashed line), the Lorc\'e-like first order current (long-dashed line) and the NKR formalism
(continuous line).}
\end{figure}
%%%%%%%%%%%%%%%%%%%%%%%%%%%%%%%%%%%%%%%%%%%%%%%%%%%%%%%%%%%%%%%%%%%%%%%%%%%%%%%%%%%%%%%%%%%%%%%%%%%
The negative parity projector is found as:
\aeq
P_{\a\b}(p)=\sum_{\l}u_\a(p,\l)\overline{u}_\b(p,\l)=\frac{\not{p}-m}{2m} P^{(3/2)}_{\a\b}(p).
\ceq
The  corresponding differential cross section of the process is then calculated as:
\aeq
\frac{d\s}{d\Omega}=\frac{1}{4}\frac{1}{(4\p)^2}\frac{\overline{\vert\M\vert^2}}{m^2}\(\frac{\o'}{\o}\)^2,
\ceq
with $\o\,(\o')$ being the energy of the incident (scattered) photon.
Performing all the calculations with the aid of the Feyncalc package, we express the final  result
in terms of the  energy variable $\h=\o/m$, and the scattering angle in the laboratory  frame $\q$
according to
\aeq\label{soflc}
\frac{d\s}{d\Omega}=\frac{r_0^2}{(1-(x-1)\h)^7}\sum_{l=0}^{10}h_l \h^l, \quad x=\cos\theta.
\ceq
Here,
\begin{subequations}
\begin{align}
h_0&=+\frac{1}{2} \left(x^2+1\right),\\
h_1&=-\frac{5}{2} (x-1) \left(x^2+1\right),\\
h_2&=+\frac{1}{162} (x (5 x (18 x (9 x-19)+581)-3852)+1815),\\
h_3&=-\frac{1}{81} (x-1) (x (5 x (9 x (9 x-22)+676)-5274)+2415),\\
h_4&=+\frac{1}{162} (x (x (x (15 x (9 x (3 x-16)+809)-35632)+51459)-33248)+16465),\\
h_5&=-\frac{1}{162} (x-1) (x (x (x (x (9 x (9 x-76)+7957)-29336)+51883)-35820)+34191),\\
h_6&=-\frac{1}{162} (x-1) (x (x (x (x (5 x (18 x-617)+17331)-46468)+57194)-61287)+39361),\\
h_7&=-\frac{1}{81} (x-1)^2 (x (x (x (x (255 x-2099)+6972)-10508)+15525)-13281),\\
h_8&=-\frac{1}{81} (x-1)^3 (x (x (x (162 x-701)+1685)-4215)+6205),\\
h_9&=-\frac{56}{81} (x-1)^4 (9 x-37),\\
h_{10}&=-\frac{392}{81} (x-1)^5.
\end{align}
\end{subequations}
We notice that the linear Lorc\'e-like current constructed within the NKR formalism  has  the correct low energy limit $(x^2+1)(r_0^2/2)$.
The result for the forward direction is then deduced from the last equation as,
\aeq
\[\frac{d\s}{d\Omega}\]_{\q=0}=\frac{1}{81}(4712\h^4-16 \h^4-16^2+81)r_0^2.
\ceq
In Fig. \ref{low} we compare the predictions of the Rarita-Schwinger--, the NKR--, and the  Lorc\'e-like currents for the
low energy differential cross section (\ref{soflc}).
We observe that all three formulations amount to equally good  and realistic predictions in the low energy regime and posses
the correct classical limit prescribed by  the Thompson cross section (dotted line in the Fig \ref{low}).
This is due to the circumstance that the Thomson limit is entirely governed my the lowest multipole,
which is the electric monopole, and is insensitive to both the spin degrees of freedom, and the boost.
Beyond this limit, we observe that the current in (\ref{Cedric_32}) gives a differential cross section which grows rapidly
with the energy in the forward direction, as shown in Fig \ref{high} for $\h=0.5$. Only the line corresponding to
the parametrization of the NKR  current leads to a differential cross section
independent of the energy in the forward direction. On this grounds we  conclude  that
linear  Lorentz invariant currents of equal electric-monopole and magnetic-dipole moments
do not necessarily lead to equivalent forward Compton scattering cross sections. A realistic description is
provided only by the current consistent with the NKR spin-3/2 Lagrangian. This finding is restricted to the NKR method alone and
does not rule out  the possibility of obtaining a correct high-energy limit of the Compton scattering cross section
within a consistent light-cone approach based on the genuine second order Lorc{\'e } current.

\end{document}